\documentclass[12pt]{article}
\usepackage{amsmath,amsfonts,amssymb}

\usepackage[sort&compress,numbers,square]{natbib}

\usepackage{mathtools}
\usepackage{hyperref}
\usepackage{placeins}
\usepackage[bbgreekl]{mathbbol}
\usepackage{bbm,dsfont}
\usepackage{graphicx}
\usepackage{graphics}
\usepackage{color}
\usepackage{epsfig}
\usepackage{psfrag}	
\usepackage{tikz}
\usetikzlibrary{snakes}
\usepackage{slashed}
\usepackage{ulem}
\usepackage[font=small,labelfont=bf,width=1\textwidth]{caption}
\usepackage{comment}
\usepackage[all]{xy}
\usepackage{multirow}
\usepackage{float}
\usepackage{tensor}
\usepackage{multirow}
\usepackage{booktabs}
\usepackage{fancyhdr}
\usepackage{upgreek}
\usepackage[T1]{fontenc}

\usepackage[none]{hyphenat}
\usepackage{color}

\newcommand \be{\begin{eqnarray}}
\newcommand \ee{\end{eqnarray}}

\numberwithin{equation}{section}

\def\RGP#1{{\color{orange} [RGP: #1]}}
\def\EC#1{{\color{red} [EC: #1]}}

\def\ie{\textit{i.e.} }
\def\eg{\textit{e.g.} }
\def\rhs{right-hand-side }
\def\lhs{left-hand-side }

\DeclareMathOperator{\Res}{Res}

\let\Re\undefined
\DeclareMathOperator{\Re}{Re}

\DeclareMathOperator{\diag}{diag}

\DeclareMathOperator{\Disc}{Disc}

\def\bC {\mathbb{C}}

\def\bR {\mathbb{R}}

\def\bZ {\mathbb{Z}}

\def\bmF\mathbbm{F}
\def\bmD\mathbbm{D}
\def\zb{{\Bar{z}}}

\newcommand{\bea}{\begin{eqnarray}}
\newcommand{\eea}{\end{eqnarray}}
\newcommand{\beq}{\begin{equation}}
\newcommand{\eeq}{\end{equation}}
\newcommand{\bal}{\begin{equation}\begin{aligned}}
\newcommand{\eal}{\end{aligned} \end{equation}}

\newcommand{\vev}[1]{{\left< {#1} \right>}}

\newcommand{\abs}[1]{{\left| {#1} \right|}}

\newcommand{\address}[1]{\vbox{\center\em#1}}
\renewcommand{\title}[1]{\vbox{\center\huge{#1}}\vspace{5mm}}

\newcommand{\cA}{{\mathcal A}}

\newcommand{\cC}{{\mathcal C}}
\newcommand{\cD}{{\mathcal D}}

\newcommand{\cM}{{\mathcal M}}
\newcommand{\cN}{{\mathcal N}}

\newcommand{\cT}{{\mathcal T}}

\newcommand{\cO}{{\mathcal O}}

\newcommand{\al}{\alpha}

\renewcommand{\sl}{\mathfrak{s}\mathfrak{l}}

\renewcommand{\b}{\beta}

\newif\ifshowexcursus

\showexcursustrue 

\definecolor{midgray}{gray}{0.4}

\begin{document}

\begin{titlepage}
\begin{center}

\vspace*{20mm}

\title{Celestial Regge theory}

\vspace{5mm}

\renewcommand{\thefootnote}{$\alph{footnote}$}

Eduardo Casali$^1$ and Riccardo Giordana Pozzi$^{1,2,3}$

\vskip 6mm

\address{
$\,^1$Departamento de Física Matemática, Instituto de Física, Universidade de São Paulo,\\
Rua do Matão 1371, São Paulo, SP 05508-090, Brazil\,,\\[0.15cm]
$\,^2$Dipartimento di Scienze Fisiche, Informatiche e Matematiche, \\
Universit\`a di Modena e Reggio Emilia, via Campi 213/A, 41125 Modena, Italy,\\[0.15cm]
$\,^3$INFN, Sezione di Bologna, via Irnerio 46, 40126 Bologna, Italy\\
}

\vskip 5mm

\tt{ecasali@usp.br, riccardo.pozzi@unimore.it}

\renewcommand{\thefootnote}{\arabic{footnote}}
\setcounter{footnote}{0}

\end{center}

\vspace{8mm}
\abstract{
\normalsize{
\noindent
Exploiting the analytic properties of scattering amplitudes, we provide an alternative but equivalent definition of the standard Mellin transform used to obtain celestial correlation functions. From this representation, we identify a celestial dispersion relation that relates the reduced correlation function to the poles and discontinuities of the bulk amplitude, {and we present a novel expansion for the celestial correlator from an integral transform of the Froissart-Gribov expansion on the bulk}. By drawing an analogy with the standard CFT case, we define the celestial Regge limit and identify the relevant celestial CFT data in terms of the partial amplitudes governing the bulk Regge limit.

}}
\vfill

\end{titlepage}
\tableofcontents

\newpage

\section{Introduction and summary}
\label{sec: Introduction and summary}

The study of holography in flat space has received new impetus in the past few years from the study of scattering amplitudes in flat space. In contrast to AdS where one can follow Maldacena's construction, flat holography space lacks a general construction from string theory, though some particular examples are known \cite{Banks:1996vh,Susskind:1997cw,Costello:2022jpg}. One way to try to make progress is to take a bottom-up approach, leveraging the vast knowledge about scattering amplitudes in flat space to try and glean some aspects of a putative holographic dual. There is now quite some literature on this approach, see \cite{Donnay:2023mrd,Raclariu:2021zjz,Pasterski:2021raf} for reviews. One of the central ideas emerging from this approach is that the holographic dual should be defined on the sphere at null infinity, where the Lorentz group $SL(2,\mathbb{C})$ acts as conformal transformations, leading naturally to the conjecture that the dual theory is some sort of CFT, dubbed celestial CFT. The relevant quantities of interest in this context are the so called celestial correlators which, beside an overall kinematical factor, are obtained by Mellin transforming the scattering amplitude. For four external massless particles, it has the expression 
\begin{equation*}
    f_{\Delta_i}(z,\bar{z})=\delta(z-\zb)(1-z)^{\frac{\Delta_{12}-\Delta_{34}}{2}}\, g(\beta,z)\,,
\end{equation*}
where $z$ is the cross-ratio, $\Delta_i$ are the external conformal dimensions, $\beta=\sum_i\Delta_i-4$, and the delta function is inherited from $4d$ bulk translational invariance. The function $ g(\beta,z)$ crucially depends on the external kinematic considered. For example, for the $s$-channel kinematics, \ie $1,2\rightarrow3,4$, has
\begin{equation*}
    g_{\Delta_i}(z,\zb)= 2^{-2-\beta}z^2\int_0^{\infty} d\omega \,\omega^{\beta-1}\cT(s,t)\Bigr\vert_{s=\omega^2,\, t=-\omega^2/z}\,.
\end{equation*}
where, $\cT(s,t)$ is the stripped scattering amplitude and $z$ is the cross ratio constrained to $z\in [1,\infty)$. See section \ref{sec: preliminaries} and appendix \ref{app: celestial 4-point function} for more details.

In  analogy to the standard discussion for scattering amplitudes in momentum space, in this work, we exploit the Cauchy theorem to rewrite the celestial amplitude as a contour integral. By proper contour modifications, we can exchanged and perform the Mellin integral. This way we find the Mellin transformed expression given by
\begin{equation*}
    g(\beta,z) =  \frac{2^{-3-\b}\,\pi}{\sin(\pi\b/2)}z^{2}\int_{\cC^{-1}(\bR^++i\varepsilon)} \frac{d\mu}{2\pi i}(-\mu+i\varepsilon)^{\b/2-1}\,\cT(s,t)\vert_{\substack{s=\mu\,\,\,\,\,\,\,\\[0.05pt] t=-\mu/z}}\,,
\end{equation*}
where $\cC^{-1}(\bR^++i\varepsilon)$ is a clockwise contour around the (shifted) positive real semi-axis. This expression holds for the $s$-channel kinematics, but similar relations can also be obtained for the other configurations. This procedure leads to a clear bridge between the analytic properties of the scattering amplitudes and those of the celestial correlators. Under suitable assumptions on the decay of the amplitudes, we obtain a dispersion relation for the celestial correlator which reads
\begin{equation*}
    g(\beta,z) =  \frac{2^{-3-\b}\,\pi}{\sin(\pi\b/2)}z^{2}\Bigl( \sum_{\mu_i}(-\mu_i)^{\b/2-1}\Res[\cT(z,\mu_i)]+\sum_{ C_i}\int_{C_i}\frac{d\mu}{\pi}(-\mu)^{\b/2-1}\,\Disc_\mu[\cT(z,\mu)]\Bigr)\,\nonumber\,.
\end{equation*}
where $\mu_i$ and $C_i$ are the amplitude's poles and cuts respectively. 

The results of this paper make use of the Froissart-Gribov expansion, which can be understood as an expansion in intermediate multi-particle states adapted to massless kinematics. For instance, focusing on the $s$-channel, the Froissart-Gribov expansion reads
\begin{equation*}
    \cT(s,t) = 2\sum_{J=0}^{\infty} a_J(s)P_J(z_s)\,.
\end{equation*}
where $z_s$ denotes the scattering angle, related to the cross-ratio by $z_s = 1 - 2/z$.

From this expansion, one can derive a corresponding representation for the celestial correlator, which is entirely determined by the partial amplitudes $a_J(s)$ and takes the form
\begin{equation*}
    f_{\Delta_i}(z,\zb)= \delta(z-\zb)(1-z)^{\frac{\Delta_{12}-\Delta_{34}}{2}}\frac{2^{-2-\b}\,\pi}{\sin(\pi\b/2)}z^{2}\sum_{J=0}^{\infty}\bar a_J(\beta) P_J\Bigl(1-\frac{2}{z}\Bigr)\,,
\end{equation*}
with
\begin{equation*}
    \bar a_J(\beta) = \int_{\cC^{-1}(\bR^++i\varepsilon)} \frac{d\mu}{2\pi i}(-\mu+i\varepsilon)^{\b/2-1}\,a_J(s)\Bigr\vert_{s=\mu}\,.
\end{equation*}
Compared to other known expansions, such as the conformal block expansion and the relativistic partial wave expansion, this relation provides a direct bulk origin for the coefficients appearing in front of the partial waves. These contributions can be interpreted as resonances, encoding the exchange of multi-particle states in the bulk amplitude. 


The Froissart-Gribov expansion is also commonly used to extract the Regge behavior of scattering amplitudes. In this case, one instead considers the $t$-channel expansion, which is, however, not well-behaved in the naive $s \to \infty$ limit, as it leads to a divergent sum over spins. To overcome this issue, one first performs an analytic continuation in the angular momentum, implemented via the Sommerfeld-Watson transform. Under suitable assumptions on the analytic structure of the partial waves, the resulting contour can then be deformed, allowing one to isolate the contribution of Regge poles, thus leading to the Regge limit of the scattering amplitude. We can then apply our contour deformations of the Mellin transform in order to define the celestial Regge limit in the bulk. 

In order to better connect to the Regge limit in CFTs, which is typically associated with small cross-ratio, in the above we consider the external kinematics of $1,3\rightarrow2,4$ where the cross ratio is constrained to lie in the interval $z\in[0,1]$ and the Regge limit is given by $z\rightarrow0 $. After an appropriate discussion we find and expression for the bulk Regge limit
\begin{equation*}
    f^{\pm}_{\Delta_i}{(z,\zb)} \approx- 2\pi \delta(\theta)\int_{-\infty}^{+\infty} d\nu\,e^{i\pi j(\nu)} \alpha^{\pm}(\nu)\,\zeta^{1-j(\nu)} \,,\nonumber
\end{equation*}
where we considered contour modifications $\mu=1+i\nu$, the cross-ratio as $z=e^{i\theta}\zeta$  taking $\zeta\ll 1$ and we collected the remaining terms in the $\al^\pm(\nu)$ coefficients by
\begin{align*}
    \alpha^{+}(\nu) &= e^{i\pi j(\nu)/2}\frac{\,2^{2 j(\nu)-\beta-7}\rho^+(\mu)\,{(1+i\nu)}^{\beta/2-1}}{\pi\,\sin(\pi\beta/2)\sin(\pi j(\nu)/2)}\,,\\[0.3cm]
    \alpha^{-}(\nu) &= e^{i\pi j(\nu)/2}\frac{\,2^{2 j(\nu)-\beta-7}\,\rho^-(\nu)\,(1 +i\nu)^{\beta/2-1}}{i\pi\,\sin(\pi\beta/2)\cos(\pi j(\nu)/2)}\,,
\end{align*}
with $\rho^\pm(\mu)$ being the residues at the Regge pole, \ie $J=j(t)$, of the analytically continued $a_J(t)$.

 We then take a page from the Regge limit of CFTs \cite{Costa:2012cb,Cornalba:2007fs,Caron-Huot:2017vep} reviewing the main features and expression. Given that this is limit emerges intrinsically from a Lorentzian, we discuss ways to rotate back to Euclidean in order to compare the results with the celestial one, thus obtaining the expression
\begin{equation*}
    \mathfrak{f}^{\pm}_{\Delta_i}(z,\zb)\approx - \,2\pi\delta(\theta)\int_{-\infty}^{+\infty} d\nu\,e^{i\pi j(\nu)} \gamma^{\pm}(\nu)\,\zeta^{1-j(\nu)}\,, \nonumber
\end{equation*}
where $\mathfrak{f}^{\pm}_{\Delta_i}(z,\zb)$ is in this case is the reduced correlator of a $D=2$ CFT, where we only used the covariance under the global $SL(2,\bC)$.  The coefficients $\gamma^\pm(\nu)$ are related to the CFT Regge pole $\sigma^\pm(\nu)$ by
\begin{align}
    \gamma^{+}(\nu) &= \frac{e^{-i\pi j(\nu)/2}\,\sigma^{+}(\nu)}{4\sin(\pi j(\nu)/2)}\frac{ e^{\pi i\frac{\Delta_{12}-\Delta_{34}}{2}}}{i \pi^2\, K_{\Delta(\nu),j(\nu)}}\,,\nonumber\\
    \gamma^{-}(\nu) &= \frac{e^{-i\pi (j(\nu)+1)/2}\,\sigma^{-}(\nu)}{4\cos(\pi j(\nu)/2)}\frac{e^{\pi i\frac{\Delta_{12}-\Delta_{34}}{2}}}{i\pi^2\, K_{\Delta(\nu),j(\nu)}}\,.\nonumber
\end{align}
For more details see section \ref{sec:CFT Regge}.
This expansion is remarkably similar to the Mellin transform of the bulk Regge limit presented above, and we can immediately read off a relation between $\textit{bulk}$ partial wave coefficients and \textit{boundary} conformal data
\begin{equation*}
    \sigma^{\pm}(\nu) = i\pi\,e^{2\pi i j(\nu)}\frac{2^{2j(\nu)-\beta-5}\Delta(\nu)^{\beta/2-1}}{\sin(\pi\beta/2)}e^{\pi i\frac{\Delta_{12}-\Delta_{34}}{2}}K_{\Delta(\nu),j(\nu)}\rho^{\pm}(\nu)\,,\nonumber
\end{equation*}
with $\Delta(\nu)=1+i\nu$.

The article is organized as follows. In section \ref{sec: preliminaries} we define conventions regarding celestial amplitudes and kinematics of the Regge limit. In section \ref{sec: Regge limit} we discuss the Regge limit of amplitudes for to the massless case. Section \ref{sec:celestial Regge} computes the Mellin transform of the Regge amplitude by introducing a contour trick and provides the celestial Froissart-Gribov expansion. Section \ref{sec:CFT Regge} reviews the CFT Regge limit and adapts it to the celestial sphere. Section \ref{sec:discussion} is a brief discussion of our results. The appendices review and expand upon results used in the main text.

\section{Preliminaries}
\label{sec: preliminaries}

We begin by recalling some properties and definitions of the four-point amplitude and its relation to the celestial correlation function.
Working on $\bR^{1,3}$, we write the four-momenta in coordinates adapted to null infinity as
\begin{equation*}
    p^\mu_i(z_i,\zb_i) = \eta_i\omega_i q_i^{\mu}(z_i,\zb_i)\,,
\end{equation*}
where $\eta_i=\pm$ labels incoming ($-$) and outgoing ($+$) particles, and $\omega_i$ is the energy. The complex variables $z_i$ are related to the azimuthal and polar angle of the stereographic projection and determine the null-momenta
\begin{equation*}
    q_i^{\mu}(z_i,\zb_i)= (1+z_i\zb_i, z_i - \zb_i, -i(z_i-\zb_i), 1-z_i\zb_i)\,.
\end{equation*}
In terms of these variables, the amplitude is written as
\begin{equation*}
    A^{(n)}(\omega_i,z_i;\ell_i)=\delta^{(4)}(\,\sum_{i=1}^{n} p_i^\mu\,)\,\cT^{(n)}(p_i^\mu(z_i,\zb_i);\ell_i)\,,
\end{equation*}
where $\cT^{(n)}(p_i^\mu(z_i,\zb_i);\ell_i)$ is the stripped $n$-point amplitude and $\ell_i$ represent the helicity of the massless external legs. 

The conformal action of the Lorentz group $SL(2,\mathbb{C})$ on the celestial sphere can be made manifest by rewriting the amplitude in terms of the boost eigenstates, which are related to the plane wave basis by a Mellin transform
\begin{equation*}
    \cA^{(n)}(z_i,\zb_i;\Delta_i,\ell_i)= \Bigl(\prod^{n}_{i=1}\int_{0}^{\infty}d\omega_i\,\omega_i^{\Delta_i-1}\Bigr)A^{(n)}(\omega_i,z_i,\zb_i; \ell_i)\,.
\end{equation*}
The Mellin transformed amplitude is covariant under the action of $SL(2,\bC)$ which, acting as global conformal transformations on the celestial sphere, it resembles a conformal correlator of a $2d$ CFT \cite{Pasterski:2016qvg,Pasterski:2017ylz}. The Mellin conjugate to the energies, \ie $\Delta_i$, can be regarded as the conformal dimensions $\Delta_i$ of the $i$-th external insertion. Together with $\ell_i$, these quantities define the (anti)holomorphic conformal weights
\begin{equation*}
    h_i= \frac{\Delta_i+\ell_i}{2} \,,\qquad \bar h_i=\frac{\Delta_i-\ell_i}{2}\,.
\end{equation*}
For the rest of the work we will be mostly interested with four-particles processes related to the four-point correlation functions by
\begin{equation*}
    \cA^{(4)}(z_i,\zb_i; \Delta_i,\ell_i)= \Bigl(\prod^{4}_{i=1}\int_{0}^{\infty}d\omega_i\,\omega_i^{\Delta_i-1}\Bigr)A^{(4)}(\omega_i,z_i,\zb_i,\ell_i)\,.
\end{equation*}
Using the full Poincaré symmetry, $\mathrm{SL}(2,\mathbb{C})\ltimes\mathbb{R}^{1,3}$, the amplitude can be put in the form
\begin{equation} \label{celestial_corr}
\cA^{(4)}(z_i,\zb_i;\Delta_i,\ell_i)=\frac{\Big(\frac{z_{14}}{z_{13}}\Big)^{h_3-h_4}\Big(\frac{z_{24}}{z_{14}}\Big)^{h_1-h_2}\Big(\frac{\bar{z}_{14}}{\bar{z}_{13}}\Big)^{\bar{h}_3-\bar{h}_4}\Big(\frac{\bar{z}_{24}}{\bar{z}_{14}}\Big)^{\bar{h}_1-\bar{h}_2}}{z_{12}^{h_1+h_2}z_{34}^{h_3+h_4}\bar{z}_{12}^{\bar{h}_1+\bar{h}_2}\bar{z}_{34}^{\bar{h}_3+\bar{h}_4}}f_{\Delta_i,\ell_i}(z,\bar{z})\,.
\end{equation}
For simplicity, we will call all the overall contribution coming from the usual $SL(2,\bC)$ covariance $ K(z_i,\zb_i,h_i,\bar h_i)$. The additional $\mathbb{R}^{1,3}$ fixes
\begin{equation}\label{reduced amplitude-s}
f_{\Delta_i,\ell_i,\eta_i}(z,\bar{z})=(z-1)^{\frac{1}{2}(h_1-h_2-h_3+h_4)}(\zb-1)^{\frac{1}{2}(\bar h_1-\bar h_2-\bar h_3+\bar h_4)}\delta(z-\zb)\,g_{\Delta_i,\ell_i,\eta_i}(z,\bar{z})\,,
\end{equation}
with the conformal cross-ratio $z$ and $\zb=z^*$ . We will refer to \eqref{celestial_corr} as the celestial correlator, as it should be the output of a putative dual theory on the celestial sphere. Compared to the typical CFT correlators, one has additional constraints coming from the momentum conservation \cite{Law:2019glh}. These lead to a distributional four point correlator with support on the real line, \ie the equator of the celestial sphere. We also note that the function depends explicitly on $\eta_i$'s, that is, on weather the momenta are incoming or outgoing. This implies that there are three disconnected regions related to the three different kinematics. We consider first the $s$-channel kinematics, $\eta_1=\eta_2=-1$ and $\eta_3=\eta_4=1$, associated to the expression \eqref{reduced amplitude-s}, where
\begin{equation*}
    s=-(p_1+p_2)^2=\omega^2\,,\qquad t=-(p_1+p_3)^2=-\omega^2/z\,,
\end{equation*}
such that $s+t+u=0$. Cross-ratios are related to the coordinates of the operator insertions by 
\begin{equation*}
    z=\frac{z_{12}z_{34}}{z_{13}z_{24}}\,\qquad \zb=\frac{\zb_{12}\zb_{34}}{\zb_{13}\zb_{24}}\,.
\end{equation*}
Note that, being related to the scattering angle, the cross-ratio can be expressed in terms of the Mandelstam variables. The domain is restricted for this configuration to
\begin{equation}
\label{eq. z = f(s,t) in the s-kinematical channel}
    z=-s/t\in [1,\infty)\subset\bR\,.
\end{equation}

The reduced correlator $f_{\Delta_i,\ell_i}(z,\zb)$ can be expressed in terms of the amplitude by picking the standard conformal frame, see appendix \ref{app: celestial 4-point function}. For the massless scalar case in the $s$-kinematics, the explicit evaluation leads to  
\begin{equation}\label{reduced amplitude}
f_{\Delta_i}(z,\bar{z})=\delta(z-\zb)(z-1)^{\frac{\Delta_{12}-\Delta_{34}}{2}}\,2^{-2-\beta}z^2\int_0^{\infty} d\omega \,\omega^{\beta-1}\cT(s,t)\Bigr\vert_{s=\omega^2,\, t=-\omega^2/z}\,,
\end{equation}
where $\beta$ is specified by the dimensions of all external particles by
\begin{equation*}
    \beta=\sum_i\Delta_i-4\,.
\end{equation*}
We further define
\begin{equation}
\label{eq reduced correlator g of the s-kinematics}
    g_{\Delta_i}(z,\zb)= 2^{-2-\beta}z^2\int_0^{\infty} d\omega \,\omega^{\beta-1}\cT(s,t)\Bigr\vert_{s=\omega^2,\, t=-\omega^2/z}\,.
\end{equation}
However, we will primarily be working in the $t$-channel kinematics, since this choice allows for a more transparent discussion of the Regge limit. In particular, we will be interested in obtaining the high-energy center of mass scattering $s\rightarrow\infty$ at $t$-fixed. As we will see, the analogous configuration of the Regge limit in CFT is associated to small cross-ratios. This configurations is excluded by the $s$-kinematics as $z$ is bounded by the kinematical sector considered. We thus perform this study in the more natural $t$-kinematics, where the limit $z\rightarrow0$ is associated to $s\rightarrow\infty$.

With this in mind, we consider the kinematics where $p_1$ and $p_3$ are incoming so $\eta_1=\eta_3=-1$, $\eta_2=\eta_4=1$, with
\begin{equation*}
    s=-(p_1+p_3)^2=\omega^2\,,\qquad t=-(p_1+p_2)^2=-z\,\omega^2\,.
\end{equation*}
In this configuration, the cross-ratio is restricted to the domain
\begin{equation*}
    z=-t/s\in [0,1]\subset\bR\,,
\end{equation*}
which is the crossed sector of the previous one.
The explicit evaluation in this channel (appendix \ref{app: celestial 4-point function}) leads to
\begin{equation}
\label{eq reduced correlator f of the t-kinematics}
    f_{\Delta_i}(z,\bar{z})=\delta(z-\zb)(1-z)^{\frac{\Delta_{12}-\Delta_{34}}{2}}\,2^{-2-\beta}z^{\frac{\beta}{2}+2}\int_0^{\infty} d\omega \,\omega^{\beta-1}\cT(s,t)\Bigr\vert_{s=\omega^2,\, t=-z\omega^2}
\end{equation}
and therefore
\begin{equation}
\label{eq reduced correlator g of the t-kinematis}
    g(z,\zb)=2^{-2-\beta}z^{\frac{\beta}{2}+2}\int_0^{\infty} d\omega \,\omega^{\beta-1}\cT(s,t)\Bigr\vert_{s=\omega^2,\, t=-z\omega^2}\,.
\end{equation}

\section{Regge limit}
\label{sec: Regge limit}
In this section we discuss the Regge limit in the case of massless external particles. Although the setup differs from the standard massive case, much of the analysis closely parallels the usual discussion~\footnote{For comprehensive treatments see \cite{Collins:1977jy,Costa:2012cb,Cornalba:2007fs,Correia:2020xtr}.}.

\begin{figure}
    \centering
    \includegraphics[width=0.85\linewidth]{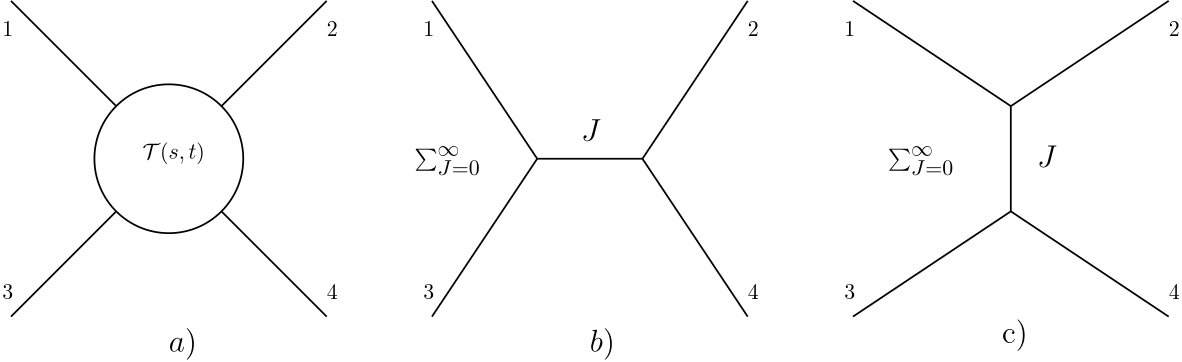}
    \caption{{Schematic representation of the Froissart-Gribov projections. A generic scattering, represented in $a)$ can be decomposed in terms of the exchange of multi-particle states of spin $J$ in the $s$- and $t$-channel, here $b)$ and $c)$ correspondingly.}}
    \label{fig:s-and-t-froissaart-gribov}
\end{figure}

We begin with the Froissart-Gribov projection of the stripped amplitude and, without loss of generality, restrict our analysis to external massless scalars. The extension to external particles carrying spin is qualitatively similar and differs mainly by the presence of additional tensor structures. The resulting expansion in $s$ reads
\begin{equation}
\label{eq Partial wave expansion s-channel}
    \cT(s,t) = 2\sum_{J=0}^{\infty} a_J(s)P_J(z_s)\,,
\end{equation}
with
\begin{equation*}
    z_s=1+2\frac{t}{s}\,.
\end{equation*}
 For the Regge limit, we will be interested in the $t$-channel expansion, which reads
\begin{equation}
\label{eq Partial wave expansion t-channel}
    \cT(s,t) = 2\sum_{J=0}^{\infty} a_J(t)P_J(z_t)\,,
\end{equation} 
where
\begin{equation*}
    z_t=1+2\frac{s}{t}\,.
\end{equation*}
We stress that the two different expansion simply gives two different ways of expanding the amplitude but they do not alter the external kinematics which is fixed, see figure \ref{fig:s-and-t-froissaart-gribov} for a schematic representation.

Even though we are considering massless particles, we retain the standard notation. Here, $J$ denotes the helicity, and $P_J(z)$ are the partial-wave eigenfunctions of the little group $SO(2)$, see appendix \ref{Appx: Details on the Froissart-Gribov expansion} for details. We refer to $a_J(s)$ and $a_J(t)$ as partial amplitudes, with the overall factor of $2$ accounting for helicity degeneracy.

The coefficients $a_J(t)$ encode the exchange of multi-particle resonance states with spin $J$. Exploiting the orthogonality of the partial waves, we can invert \eqref{eq Partial wave expansion t-channel} to obtain
\begin{equation}
\label{eq partial wave coefficients}
    a_J(t) = \frac{1}{2\cN_J}\int_{-1}^{1}dz_{t}\,\frac{\cT(s(z_t,t),t)P_J(z_t)}{\sqrt{1-z_t^2}}\,,
\end{equation}
where $\cN_J$ is a normalization factor, $s$ is written explicitly in terms of 
\begin{equation*}
    s(z_t,t)=\frac{z_t-1}{2}t
\end{equation*}
and the integral is taken over the physical region of the $t$-channel, where $s<0$ and $t>0$ and $z_t\in[-1,1]$. So, even though we are interested in the Regge limit, defined by $s \to \infty$ with $t$ fixed ($t<0$), the computation of $a_J(t)$ requires considering the amplitude in the physical $t$-channel.
Taking $z_t$ and $t$ as independent variables, the last Mandelstam variable can be written as 
\begin{equation}
\label{eq relation between u and s}
    u(s,t)=-s-t=-\frac{z_t+1}{2}t=s(-z_t,t)\,.
\end{equation}

\begin{figure}
    \centering
    \includegraphics[width=0.45\textwidth]{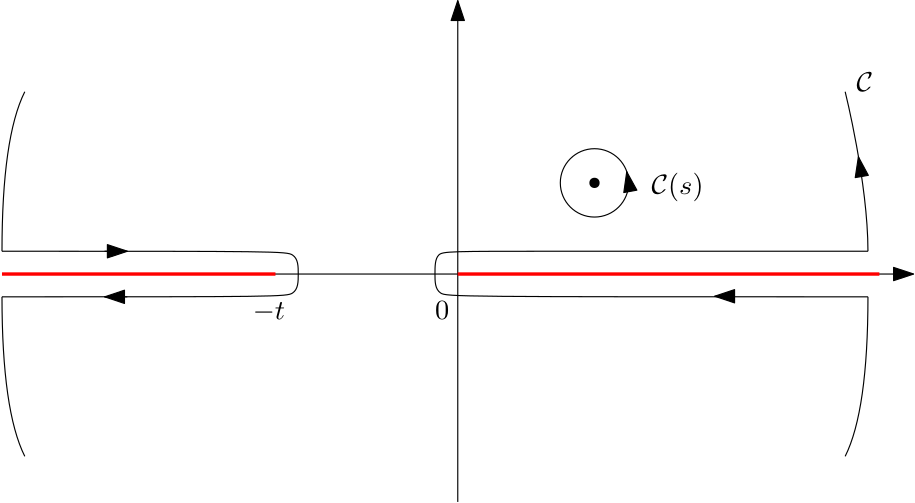}
    \caption{Contour modification from $\cC(s)$ to $\cC$ in the complex $s$-plane. Setup with no massive interactions and branch cuts are in red.}
    \label{fig:cuts}
\end{figure}

We now write the partial amplitude \eqref{eq partial wave coefficients} in such a way that the contributions coming from tree-level physics and from resonances are made explicit. For that, we 
can use Cauchy's theorem to write the amplitude as
\begin{equation*}
    \cT(s,t)=\oint_{\cC(s)}\frac{ds'}{2\pi i}\frac{\cT(s',t)}{s'-s}\,,
\end{equation*}
where the contour $\cC(s)$ encircle counterclockwise $s$. By modifying the contour, we can rewrite it as
\begin{equation}
\label{eq amplitude as pole plus C contour}
    \cT(s,t) = \sum_{i=1}^{M}\left(\frac{r_s(t)}{m_i^2-s}+\frac{r_u(t)}{m_i^2-u}\right)+\int_{\cC}\frac{ds'}{2\pi i}\frac{\cT(s',t)}{s'-s}\,.
\end{equation}
Here, we assumed $M$ massive exchanges, \eg through some $\sum_i\phi^2\Phi_i$ interactions. In order to keep the expression as general as possible we also consider the residues of such exchange $r_s(t)$  and $r_u(t)$ to be potentially a function of $t$. We also take $\cC(s)$ to be a counterclockwise contour sufficiently close to the  value of the given Mandelstam variable.
Assuming that the amplitude at infinity vanishes as
\begin{equation}
\label{eq convergence requirement for the amplitude}
    \vert \cT(s,t) \vert\sim\vert s\vert^{-a}\quad\text{for}\quad s\rightarrow\infty\,,
\end{equation}
with some $a>0$, we can discard the contributions of the arcs at infinity, figure \ref{fig:cuts}, leading to the expression 
\begin{equation}
\label{eq amplitude as poles plus disc - non-cross-sym}
    \cT(s,t) = \sum_{i=1}^{M}\left(\frac{r_s(t)}{m_i^2-s}+\frac{r_u(t)}{m_i^2-u}\right)+\frac{1}{\pi}\int_{0}^{+\infty}ds' \frac{D_s(s',t)}{s'-s}+\frac{1}{\pi}\int_{-\infty}^{-t}ds' \frac{D_s(s',t)}{s'-s}\,.
\end{equation}
The discontinuity of the amplitude is given by
\begin{equation*}
    D_s(s,t) = \frac{1}{2i}\Bigl(\cT(s(1+i\epsilon_0)+i\epsilon(1-i\epsilon_0),t,u)-\cT(s(1+i\epsilon_0)-i\epsilon(1-i\epsilon_0),t,u)\Bigr)\label{eqdisc in s}\,,
\end{equation*}
where we have considered branch cuts that are slightly tilted by a small parameter $\epsilon_0$ in order to avoid the overlap with the on-shell poles. By setting $\epsilon_0=0$ we obtain the standard expression for the discontinuity on the real axis, figure \ref{fig:cuts}.
For crossing symmetric amplitudes, one can relate by crossing the discontinuity of the $u$-cut with the discontinuity of the crossed amplitude  by
\begin{equation*}
    D_s(s,t)= - D_u(u,t)\,.
\end{equation*}
This allows us to rewrite the amplitude as
\begin{equation}
\label{eq amplitude as poles plus disc}
    \cT(s,t) = \sum_{i=1}^{M}\left(\frac{r_s(t)}{m_i^2-s}+\frac{r_u(t)}{m_i^2-u}\right)+\frac{1}{\pi}\int_{s_T}^{\infty}ds' \frac{D_s(s',t)}{s'-s}+\frac{1}{\pi}\int_{u_T}^{\infty}du' \frac{D_u(u',t)}{u'-u}\,,
\end{equation}
where $s_T$ and $u_T$ are the branch-point singularities associated with the thresholds in the $s$- and $u$-channels. Differently from the standard massive case, the threshold point, and hence the branch cuts in the complex $s$-plane, originate exactly at $s = 0$, \ie there is no mass gap.

Another remark concerns the convergence requirement in \eqref{eq convergence requirement for the amplitude}. There is a loophole that allows one to extend the discussion even when \eqref{eq convergence requirement for the amplitude} is not satisfied. In particular, if the amplitude grows as $|\cT(s,t)|\sim |s|^{N-a}$, one can define
\begin{equation*}
    \tilde\cT(s,t)=\prod^{N}_{i=1}(s-s_i)^{-1}\cT(s,t)\,,
\end{equation*}
introducing as many poles ${s}_{i=1,\dots,N}$ as required to ensure convergence, so that $|\tilde\cT(s,t)|\sim |s|^{-a}$; see \cite{Collins:1977jy}. This procedure, however, introduces $N$ unfixed constants that must be fixed by other means. We will not need this construction for our current applications, but it may prove useful in other contexts. In particular, exploiting this expression, the discussion on the next chapter can be suitably extended as needed when dealing with other cases.

With the expression for the amplitude \eqref{eq amplitude as poles plus disc} in hand, we proceed by substituting it in the definition of the partial wave coefficients \eqref{eq partial wave coefficients}. After some manipulations, see appendix \ref{app: details on the u- dicontinuity} for more details, we get
\begin{align}
    a_J(t)&=\sum_{i=1}^{M}\Bigl(\frac{1}{\cN_J}\frac{2 \,r_s^{i}(t)}{t}\frac{Q_J(z_0^{i})}{(1-z_0^i)^{1/2}}+\frac{1}{\cN_J}\frac{2 \,r_u^{i}(t)}{t}\frac{Q_J(-z_0^{i})}{(1-z_0^i)^{1/2}}\Bigr)+\nonumber\\
    &+\frac{1}{\pi\cN_J}\int_1^{\infty} dz_t'\frac{D_s(s(z_t',t),t)Q_J(z_t')}{\sqrt{z_t'^2-1}} + \frac{1}{\pi\cN_J}\int_1^{\infty} dz_t'\frac{D_u(u(-z_t',t),t)Q_J(-z_t')}{\sqrt{z_t'^2-1}}\label{eq partial amplitude explicit}\,,
\end{align}
with $z_0^{i}=1+2m_i/t$ and $Q_J(z)$ given by \eqref{eq QJ(z) explicit}
\begin{equation*}
    Q_J(z) = \frac{\pi}{2^{1+J}\,z^{J}}\,{}_2 F_1\Bigl(\frac{J}{2},\frac{1+J}{2},1+J;\frac{1}{z^2}\Bigr) = \frac{\pi}{2}\Bigl(1+\sqrt{1-\frac{1}{z^2}}\Bigr)^{-J} z^{-J}\,.
\end{equation*}
See the appendix \ref{Appx: Details on the Froissart-Gribov expansion} for more details.
Using parity, $Q_J(-z) = (-1)^J Q_J(z)$, and \eqref{eq relation between u and s} which implies\footnote{ Note that this simply follows from $2i \,D_u(u(-z_t,t),t)= \cT(u(-z_t,t)+i\epsilon ,t)-\cT(u(-z_t,t)-i\epsilon ,t) =\cT(s(z_t,t)+i\epsilon ,t)-\cT(s(z_t,t)-i\epsilon ,t)=2i\, D_s(s(z_t,t),t) $}
\begin{equation*}
    D_u(u(-z_t,t),t)=D_s(s(z_t,t),t)\,,
\end{equation*}
one sees that the contribution from the discontinuity vanishes for odd $J$
\begin{equation}
\label{eq partial amplitude explicit even}
    a_J(t)=0\,,
\end{equation}
while for even $J$ one has
\begin{equation}
\label{eq partial amplitude explicit odd}
    a_J(t)=\sum_{i=1}^{M}\Bigl(\frac{1}{\cN_J}\frac{2 \,r_s^{i}(t)}{t}\frac{Q_J(z_0^{i})}{(1-z_0^i)^{1/2}}+\frac{1}{\cN_J}\frac{2 \,r_u^{i}(t)}{t}\frac{Q_J(-z_0^{i})}{(1-z_0^i)^{1/2}}\Bigr)+\frac{2}{\pi\cN_J}\int_1^{\infty} dz_t'\frac{D_s(s(z_t',t),t)Q_J(z_t')}{\sqrt{z_t'^2-1}}\,.
\end{equation}

The fact that only even spins receive a contribution from the discontinuity is a consequence of crossing symmetry, which allows to exclude the odd spin partial amplitudes. This can be easily seen by expanding both sides of the crossing equation
\begin{equation*}
    \cT(s,t)=\cT(u,t)
\end{equation*}
a la Froissart-Gribov, while keeping the $t$-channel variable fixed. For $\cT(u,t)$ this is
\begin{equation}
\label{eq FG expansion with the u}
    \cT(u,t) = 2\sum a_J(t) P_J(-z_t)
\end{equation}
which follows from $s(-z_t,t)=u(z_t,t)$. We thus have
\begin{equation*}
    \cT(u,t)=\cT(s(-z_t,t),t)
\end{equation*}
giving \eqref{eq FG expansion with the u}. Crossing is therefore simply given by
\begin{equation*}
    \cT(z_t,t)=\cT(-z_t,t)\,.
\end{equation*}
Combining this with parity properties of the partial waves, \ie $P_J(-z)=(-1)^J P_J(z)$, one sees that only even spins partial amplitudes are allowed.

We now focus on the kinematic regime of large $s$ at fixed $t$, corresponding to the Regge limit. In this regime, however, the expansion \eqref{eq Partial wave expansion t-channel} ceases to converge, since for large $z_t$, or equivalently large $s$, the partial waves scale as 
\begin{equation*}
    P_J(z_t) \sim \frac{z_t^J}{2^{1-J}}\,.
\end{equation*}
The standard way to circumvent this issue is to perform a Sommerfeld-Watson transform, which allows one to replace the explicit sum in \eqref{eq Partial wave expansion t-channel} with a complex integral over the analytically continued spin $J$. We treat even and odd spins separately by introducing partial amplitudes with definite signature, so as to better organize the contributions arising from the discontinuities.
Besides making crossing properties manifest, this procedure improves the convergence of the integral when large spins are involved, especially for non-crossing-symmetric amplitudes.
Indeed, from \eqref{eq partial amplitude explicit} one finds that the two integrals behave rather differently upon analytically continuing $J$ to complex values, due to the analytic structure of $Q_J(z)$. Working with definite-signature amplitudes allows these contributions to be properly organized and combined so as to improve the convergence, as reviewed in appendix \ref{app: Details on the large spin partial amplitudes behavior}.

Using \eqref{eq partial amplitude explicit even} and \eqref{eq partial amplitude explicit odd} we define the definite spin contributions by
\begin{equation*}
    a_{J}^+(t) = \sum_{i=1}^{M}\frac{1}{\cN_J}\frac{2 \,r^{i\,+}(t)}{t}\frac{Q_J(z_0)}{(1-z_0^i)^{1/2}}+ \frac{2}{\cN_J}\int_1^{\infty} dz_t'\frac{D_s(s(z_t',t),t)Q_J(z_t')}{\sqrt{z_t'^2-1}}\nonumber\,,\\
\end{equation*}
where $a_J^{+}(t)$ stands for even spins and $r^{i\,+}(t) = r_s^i(t)+r_u^i(t)$. These definitions can be exploited to split the amplitude into definite signature amplitudes
\begin{equation*}
    \cT(s,t)=\cT^{+}(s,t)+\cT^{-}(s,t)\,,
\end{equation*}
which will be therefore have the corresponding Froissart-Gribov expansion 
\begin{equation*}
    \cT^{\pm}(s,t)= \sum_{J=0}^{\infty}a_{J}^{\pm}(t)(P_J(z_t)\pm P_J(-z_t))\,.
\end{equation*}
It is on them that we perform the Sommerfeld-Watson transform which leads to
\begin{equation*}
    \cT^{\pm}(s,t) = \int_{\cC}\, \frac{dJ}{2i\, \sin(\pi J)} a^{\pm}_{J}(t)\Bigl(P_J(z_t)\pm P_J(-z_t)\Bigr)\,,
\end{equation*}
where $\cC$ goes around the positive real axis so that the evaluation of the residues of $1/\sin(\pi J)$ leads back to the sum above. 

For crossing-symmetric amplitudes only the even-spin contributions need to be considered. Nevertheless, in order to keep the discussion general, we will retain both even and odd spins, see appendix \ref{app: Details on the large spin partial amplitudes behavior}.

\begin{figure}
    \centering
    \includegraphics[width=0.4\linewidth]{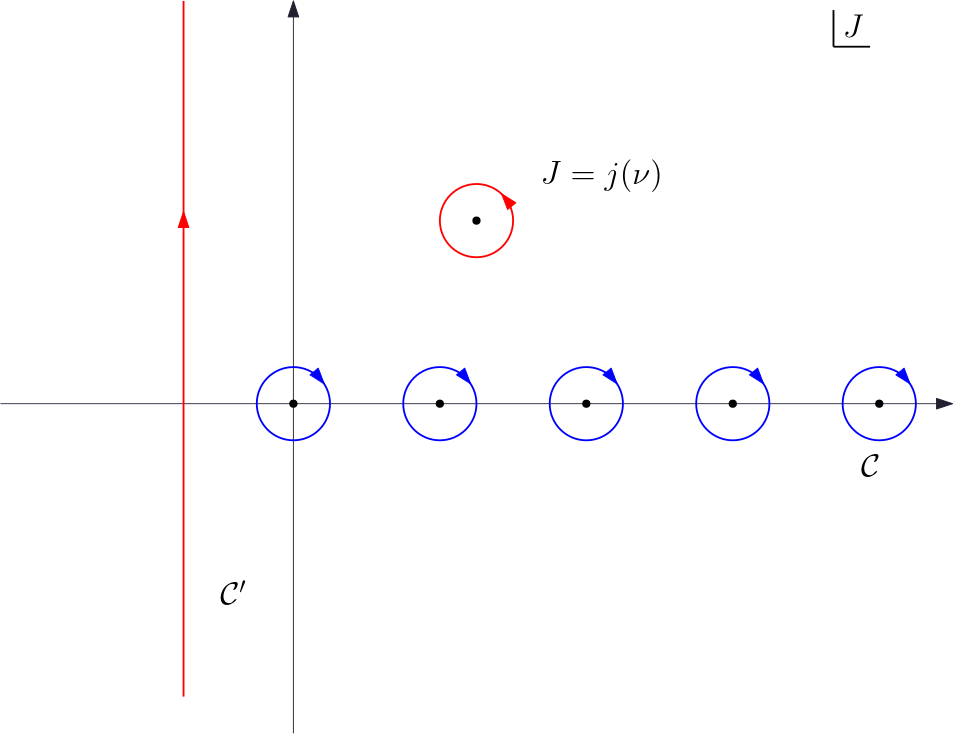}
    \caption{ A simplified illustration of the contour modification from $C$ to $C'$ giving the analytic continuation on the spin with $J=j(\nu)$ the Regge pole.}
    \label{fig: analyticity-in-J}
\end{figure}

The final assumption leading to the expected Regge behavior is that $a^{\pm}_J(t)$ are analytic throughout the complex $J$-plane, up to isolated singularities. Under this assumption, the contour $\cC$ can be deformed to pick up the relevant residues, leaving a line integral, as discussed in appendix \ref{app: regge pole} and schematically depicted in figure \ref{fig: analyticity-in-J}. We therefore take
\begin{equation*}
    a^{\pm}_J(t) = \frac{\rho^{\pm}(t)}{J-j(t)}\,,
\end{equation*}
where $\rho^\pm(t)$ are the theory-dependent residues. By contour modification we get
\begin{equation}
\label{eq. leading regge pole plus contour}
    \cT^{\pm}(s,t) = -\pi\,\rho^{\pm}(t) \frac{P_{j(t)}(z_t)\pm P_{j(t)}(-z_t)}{\sin(\pi j(t))}+\int_{\cC'}\, \frac{dJ}{2i \sin(\pi J)} a^{\pm}_{J}(t)\Bigl(P_J(z_t)\pm P_J(-z_t)\Bigr)\,.
\end{equation}

By taking $s \gg 1$, which in turn implies $|z_t| \gg 1$, we see that, in general, the first term dominates because the contour $\mathcal{C}'$ is located to the left of the Regge pole and so $\max_{\mathcal{C}'}\Re(J) < \Re(j(t))$. Consequently, the background integral is subleading, and the expression can be approximated by the residue of the leading Regge pole as

\begin{equation*}
    \cT^{\pm}(s(z_t,t),t) \approx -\pi\,\rho^{\pm}(t) \frac{z_t^{j(t)}\pm (-z_t)^{{j(t)}}}{2^{-j(t)+1}\sin(\pi j(t))}\,.
\end{equation*}

As it will be useful in the following, we already notice that the amplitude depends explicitly only in the transverse momentum $t$ and on the $t$-channel scattering angle. This expression can be thus written explicitly in terms of the ratio $z=-t/s$, which we already identified as the celestial cross-ratio and related to the $z_t$ angle by
\begin{equation*}
     z_t=1-\frac{2}{z}\,.
\end{equation*}
We can thus re-express the Regge limit in terms of $z\rightarrow0$, thus obtaining
\begin{equation}
\label{eq Regge limit of T}
    \cT^{\pm}(z,t) \approx -\pi\,\rho^{\pm}(t)(-1)^{j(t)}\frac{z^{-j(t)}}{2^{1-2j(t)}} \frac{(1\pm (-1)^{{j(t)}})}{\sin(\pi j(t))}\,.
\end{equation}

The purpose of the following section is to translate this result into the language of celestial CFT. As reviewed in the previous section, this is achieved by performing a Mellin transform. However, a direct implementation of \eqref{eq reduced correlator g of the t-kinematis} is not feasible without an explicit expression for the partial amplitudes.
In the next section, we therefore introduce an alternative--yet equivalent--procedure to carry out the Mellin transform while keeping all relevant quantities implicit.

\section{Celestial Regge theory}
\label{sec:celestial Regge}

In this section, we introduce a notion of the Regge limit on the celestial sphere, inspired by the standard Regge limit in conformal field theories \cite{Cornalba:2007fs,Costa:2012cb,Kravchuk:2018htv}. To this end, we begin by introducing an alternative formulation of the Mellin transform, assuming only generic analytic properties of the amplitude. The advantage of this approach is that it provides a more direct connection to the Regge limit of ordinary CFTs, as we will see shortly.

\subsection{Contour Trick}
\label{Contour Trick}
Let us first start with a generic example where we take the definition of the Mellin transform of a given function $f(z)$ as
\begin{equation*}
    \cM[f](s) = \int_{0}^\infty dz\, z^{s-1} f(z) \,.
\end{equation*}
We now rewrite the function $f(z)$ by exploiting Cauchy's theorem
\begin{equation*}
    f(z)=\oint_{\cC(z)}\frac{dr}{2\pi i}\frac{f(r)}{r-z}\,,
\end{equation*}
where $\cC(z)$ is a contour centered in $z$ and, substituting this expression in the definition of the Mellin transform, we simply get
\begin{equation*}
    \cM[f](s) = \int_{0}^\infty dz \oint_{\cC(z)}\frac{dr}{2\pi i}\,z^{s-1}\frac{f(r)}{r-z}\,.
\end{equation*}
This procedure has the nice effect of removing the $z$ dependence from the function, which now depends on this auxiliary complex variable $r$. This would in principle allow for a straightforward integration on $z$. However, in order to actually perform such integral, we must remove the contour dependence on $z$, so that the two integral can be swapped.

At this stage we assume $f(z)$ to be analytic with no poles or branch cuts on the domain of integration. However, if that is not the case, one can just shift those contributions by a small $i\varepsilon$, as we will see later in the section. 
Under this assumption, we modify the contour by pushing it to $\cC(z)\rightarrow\cC(\bR^+)$ where $\cC(\bR^+)$ is a counterclockwise contour that encircles the positive real axis. In general we may assume it to be sufficiently close to the axis so to exclude any feature on the complex plane coming from $f(z)$, \eg branch cuts or poles. We are therefore effectively surrounding all the integration domain of the Mellin transform.
This allow us to swap and perform the integral in $z$, leading to

\begin{equation*}
    \cM[f](s)= -\frac{\pi}{\sin(\pi s)}\int_{\cC(\bR^+)} \frac{dr}{2\pi i}(-r)^{s-1}f(r)\,.
\end{equation*}
As an example and check of this equivalent expression consider the function $f(z)=e^{-z}$. Using the original definition this gives the Gamma function $\cM[e^{-z}](s) = \Gamma(s)$. In terms of the contour integral we thus find the identity
\begin{equation*}
   - \frac{\pi}{\sin(\pi s)}\int_{\cC(\bR^+)} \frac{dr}{2\pi i}(-r)^{s-1} e^{-r} = \Gamma(s)\,.
\end{equation*}
This expression provides an alternative integral representation of the Gamma function, with the contour $\mathcal{C}(\mathbb{R}^+)$ known in the literature as the Hankel contour.

This formulation of the Mellin transform is particularly useful, as it allows one to establish a direct connection between the analytic properties of the function to be transformed and its Mellin-transformed expression. The latter can then be evaluated by exploiting the analytic features of the function and deforming the contour to pick up residues or discontinuities.

This manipulation can be applied right-away to the Mellin transform defined in \eqref{eq reduced correlator g of the t-kinematis}. In order to show the interesting features that emerge from this analysis, we present in this subsection the discussion for both  $s$ and $t$ external kinematics, even though only the latter one will be necessary for the next section.

To match the standard literature, we first start with the external $s$-kinematic where $s=-(p_1+p_2)^2$ and $t=-(p_1-p_3)^2$. In this case, the Mellin transform can be rewritten as \eqref{eq reduced correlator g of the s-kinematics}, which we repeat here for clarity
\begin{equation*}
    g(\beta,z) = 2^{-2-\b}z^{2}\int_0^{\infty}d\omega\,\omega^{\beta-1}\cT(s,t)\Bigl\vert_{\substack{s=\omega^2\\[0.05pt] t=-\omega^2/z}}\,.
\end{equation*}
We also recall that for this kinematics we have $z \in [1,\infty) \subset \mathbb{R}$. We can then trade one of the Mandelstam variables for the scattering angle or, analogously, for the celestial cross-ratio. This gives us two possible choices, depending on which channel we wish to keep explicit. Let us, for instance, take $s$ and $z$ as independent variables, with $t = t(z, s)$ and $u = -s - t$.
Following the discussion above, we start from the integral 
\begin{equation*}
    \cT(s,t)= \oint_{\cC(s+i\varepsilon)} \frac{d\mu}{2\pi i}\frac{\cT(\mu, t(z,\mu))}{\mu-s-i\varepsilon}\,,
\end{equation*}
where we have made explicit the dependence of $t$ in the other variables and we consider a counterclockwise contour $\cC(s+i\varepsilon)$ around $\mu=s+i\epsilon$. For compactness we define
\begin{equation*}
    \cT(\mu,z)=\cT(s,t)\vert_{\substack{s=\mu\,\,\,\,\,\,\,\\[0.05pt] t=t(z,\mu)}}\,.
\end{equation*}

Given that typically we expect the scattering amplitude to have a non-trivial analytic structure, as anticipated above, we have also introduced a regulator with $\varepsilon>0$ to separate in a clearer way the contributions that have different origins.

We can now take the Mellin transform   
\begin{equation}
\label{eq contour trick on the amplitude}
    \int_0^{\infty}d\omega\,\omega^{\beta-1}\oint_{\cC(s+i\varepsilon)} \frac{d\mu}{2\pi i}\frac{\cT(\mu,z)}{\mu-s-i\varepsilon}\Bigl\vert_{s=\omega^2}= \int_{\cC(\bR^++i\varepsilon)} \frac{d\mu}{2\pi i}\cT(\mu,z) \int_0^{\infty}d\omega\frac{\omega^{\beta-1}}{\mu-\omega^2-i\varepsilon}\,,
\end{equation}
where again the contour surrounds fully the domain of integration, which in this case is the slightly shifted positive real line. Here $\cC(\bR^++i\varepsilon)$ must be considered as a counterclockwise contour sufficiently close to the given line. 
Note that in the \rhs of \eqref{eq contour trick on the amplitude} the Mellin transform actually acts on a contribution that resemble that of a tree-level massive $s$-channel exchange.

Evaluating the $\omega$-integral one finds
\begin{equation*}
    \int_0^{\infty}d\omega\,\omega^{\beta-1}\cT(s,t)\Bigl\vert_{\substack{s=\omega^2\\[0.05pt] t=-z\omega^2}} = \int_{\cC(\bR^++i\varepsilon)} \frac{d\mu}{2\pi i}\cT(\mu,z)\Bigl(\frac{-\pi (-\mu+i\varepsilon)^{\b/2-1}}{2\sin(\pi\b/2)}\Bigr)\,.
\end{equation*}
The reduced correlator is then written as 
\begin{equation}
\label{eq s-channel contour mellin transform}
    g(\beta,z) =  \frac{2^{-3-\b}\,\pi}{\sin(\pi\b/2)}z^{2}\int_{\cC^{-1}(\bR^++i\varepsilon)} \frac{d\mu}{2\pi i}(-\mu+i\varepsilon)^{\b/2-1}\,\cT(\mu,z)\,,
\end{equation}
where we absorbed a sign in the orientation of the contour, which we called $\cC^{-1}$. Even though a contour integral formally remains, the Mellin transform has effectively been performed. Thus, from \eqref{eq s-channel contour mellin transform} one obtains, {\it implicitly~}\footnote{By "implicitly" we mean that, under our assumptions, this expression is valid for any amplitude considered as input and directly yields the Mellin-transformed correlation function. In contrast, the standard Mellin transform requires performing the integral explicitly and therefore an explicit expression for the amplitude is required. } the Mellin-transformed correlation function. From this expression, one can compute the reduced correlator by contour modifications.

As an example we can consider the simple case of a cubic interaction with the exchange of a particle with mass $m$ in the $s$-channel. The tree-level four-point scattering amplitude is given by
\begin{equation*}
    \cT =-g^2\frac{1}{s-m^2+i\epsilon}\,,
\end{equation*}
which has a pole at $s=m^2-i\epsilon$. We can now use this as an input for \eqref{eq s-channel contour mellin transform} giving
\begin{equation*}
    g(\beta,z) = \frac{2^{-3-\b}\,\pi}{\sin(\pi\b/2)}z^{2}\int_{\cC^{-1}(\bR^++i\varepsilon)} \frac{d\mu}{2\pi i}(-\mu+i\varepsilon)^{\b/2-1}\,\Bigl(\frac{-g^2}{\mu-m^2+i\epsilon}\Bigr)\,.
\end{equation*}
\begin{figure}
    \centering
    \includegraphics[width=0.45\linewidth]{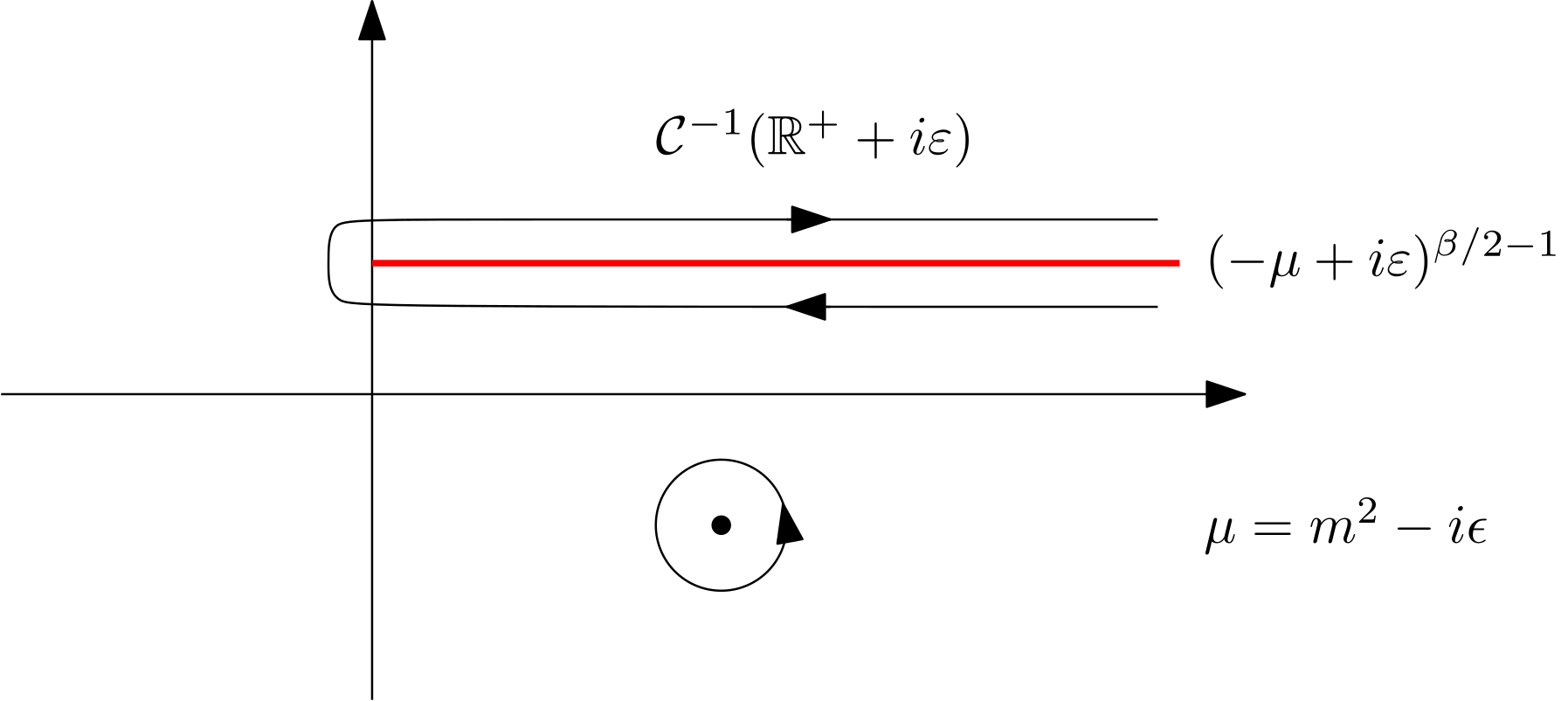}
    \caption{Contour modification for the modified Mellin transform. We move from the contour $\cC^{-1}(\bR^++i\varepsilon)$ to the one around the pole at $\mu=m^2-i\epsilon$}
    \label{fig:contour-trick}
\end{figure}
By contour modification, figure \ref{fig:contour-trick}, we can now take the residue and send the regulators to zero, leading to the known result 
\begin{equation*}
      g(\beta,z) =  \frac{2^{-3-\b}\,\pi}{\sin(\pi\b/2)}z^{2}\,g^2\,m^{\beta-2}e^{\pi i\beta/2}\,.
\end{equation*}
This example illustrates why we began our discussion by introducing the regulator. Without it, there would be an ambiguity as to whether the pole lies inside or outside the contour. These regulator helps in separating the different contributions in a way that the pole of the amplitude does not lie inside the region of the contour $\cC^{-1}(\bR^++i\varepsilon)$. Thus, in general, we can drop the regulator if we take the poles or discontinuity of the amplitude to be on the complementary region of enclose by $\cC^{-1}(\bR^++i\varepsilon)$.

Lastly, we note that this procedure also leads naturally to a celestial dispersion relation.
In particular, from \eqref{eq s-channel contour mellin transform}, we can take contour modifications which, under the assumption of vanishing arcs, leads to the expression
\begin{equation}
\label{eq celestial dispersion relation}
    g(\beta,z) =  \frac{2^{-3-\b}\,\pi}{\sin(\pi\b/2)}z^{2}\Bigl( \sum_{\mu_i}(-\mu_i)^{\b/2-1}\,\Res[\cT(\mu_i,z)]+\sum_{ C_i}\int_{C_i}\frac{d\mu}{\pi}(-\mu)^{\b/2-1}\,\Disc_\mu[\cT(\mu,z)]\Bigr)\,,
\end{equation}

\noindent giving a dispersion relation for the celestial correlator. In this expression we generically assumed some poles at $\mu_i$ and branch cut $C_i$ contributions. Notice that on the second contribution the $(-\mu)^{\b/2-1}$ lies outside the discontinuity as a consequence of the contour modification. This expression tells us that the full celestial correlator can be recovered from the knowledge of the analytic structure of the bulk amplitude only. Other dispersion relations have appeared in the literature \cite{Chang:2021wvv} which makes use of the bulk optical theorem.

Given the close similarity with the standard computation of the Mellin transform by writing it as an integral over the whole real domain, we briefly comment on the main differences between our definition and this procedure. The parity of 
 \begin{equation*}
     \cT(s,t)\Bigl\vert_{\substack{s=\omega^2\\[0.05pt] t=-z\omega^2}}=\cT(z,\omega^2)=\cT(z,(-\omega)^2)\,,
 \end{equation*}
can be exploited to trade the line integral with a contour integral. The reasoning goes as follows: start by considering the Mellin transform
 \begin{equation*}
     \int_{-\infty}^{+\infty}d\omega \omega^{\beta-1}\cT(z,\omega^2)\,.
 \end{equation*}
 One can split the integral in the positive and negative regions to obtain
 \begin{equation*}
     \int_{-\infty}^{+\infty}d\omega \omega^{\beta-1}\cT(z,\omega^2)=\int_{0}^{+\infty}d\omega \omega^{\beta-1}\cT(z,\omega^2)+\int_{-\infty}^{0}d\omega \omega^{\beta-1}\cT(z,\omega^2)\,.
 \end{equation*}
 Sending $\omega\rightarrow-\omega$ in the second contribution gives
 \begin{align*}
     \int_{-\infty}^{+\infty}d\omega \omega^{\beta-1}\cT(z,\omega^2)&=\int_{0}^{+\infty}d\omega \omega^{\beta-1}\cT(z,\omega^2)-\int_{+\infty}^{0}d\omega (-\omega)^{\beta-1}\cT(z,\omega^2)\\
     &=(1-(-1)^\beta)\int_{0}^{+\infty}d\omega \omega^{\beta-1}\cT(z,\omega^2)\\
     &=-2i e^{\pi i\beta/2}\sin(\pi\beta/2)\int_{0}^{+\infty}d\omega \omega^{\beta-1}\cT(z,\omega^2)
 \end{align*}
 One can therefore exploit the integration over the full real line by closing the contour and picking poles and discontinuities, assuming the arc at infinity does not contribute. However, the crucial difference with the contour trick discussed above is that this expression must be still evaluated. It is just a different way of writing the Mellin transform while in our case the Mellin integral has been evaluated and the leftover contour integral is crucially \textit{not} a Mellin integral. The advantage of the contour trick is that it allows to perform the Mellin transform while keeping the expressions implicit. The integral in the complex $\mu$ that we obtained by using the contour trick actually encodes, in a sense, the dynamical information, as showed also by \eqref{eq celestial dispersion relation}.

Another advantage of using \eqref{eq s-channel contour mellin transform} is that it allows us to discuss the analytic properties of the celestial amplitude by relying on those of the bulk amplitude. This is typically not the case when we express the amplitude in terms of $\omega$ in the standard Mellin transform, which leads then to discuss the analytic properties of the amplitude in the complex $\omega$ space. Of course the discussion would still be consistent with the analytic properties of the bulk amplitude, however, the identification is not as straightforward  since it appears on all the Mandelstam variables quadratically, \eg $s=\omega^2$ and $t=-z\omega^2$.

We conclude by observing that through \eqref{eq s-channel contour mellin transform} one obtains a natural prescription to implement a cutoff and probe different energy regimes. Indeed, the expression \eqref{eq s-channel contour mellin transform} already defines the Mellin-transformed correlator, and different kinematic regimes can be probed through appropriate contour deformations. Rather than introducing an explicit cutoff on the Mellin transform, for instance
\begin{equation*}
   \int_0^{\infty} d\omega \,\omega^{\beta-1}\, \cT(\omega^2,-z\omega^2) \sim \int_0^{\Lambda_{co}} d\omega \,\omega^{\beta-1}\, \cT(\omega^2,-z\omega^2)
\end{equation*} 
one can modify the contour in \eqref{eq s-channel contour mellin transform} instead.

\subsection{Froissart-Gribov expansion in the celestial sphere}
We now use the contour prescription to translate the partial wave expansions \eqref{eq Partial wave expansion s-channel} and \eqref{eq Partial wave expansion t-channel} to the celestial sphere. Here we go back to consider the external scattering kinematics as specified by the setup $1,3\rightarrow2,4$ with $s=-(p_1+p_3)^2$ and $t=-(p_1+p_2)^2$.
In this regime, where $z\in [0,1]$, the reduced correlator is specified by \eqref{eq reduced correlator g of the t-kinematis}, which we write here again for clarity
\begin{equation*}
    g(\beta,z) = 2^{-2-\b}z^{\frac{\beta}{2}+2}\int_0^{\infty}d\omega\,\omega^{\beta-1}\cT(s,t)\Bigl\vert_{\substack{s=\omega^2\\[0.05pt] t=-z\omega^2}}\,.
\end{equation*}
The discussion in the previous section goes essentially unchanged. Indeed, beside the different overall kinematical factor, one can consider the Cauchy theorem for the $s$-variable and take the Mellin integral.
However when discussing the Regge limit, as discussed in section \ref{sec: Regge limit}, we will instead trade the dependence on $s$ for $s(z_t,t)$ or, in other terms, $s=s(z,t)$ with $z$ the cross-ratio of the CCFT. In this context, we will be therefore interested in considering the contour trick applied to the $t$ dependence. So, differently than before, we consider 
\begin{equation*}
    \cT(s(z,t),t)= \oint_{\cC(t+i\varepsilon)} \frac{d\mu}{2\pi i}\frac{\cT(s(z,\mu),\mu)}{\mu-t-i\varepsilon}\,.
\end{equation*}
with the counterclockwise contour $\cC(t+i\varepsilon)$ around $\mu=t+i\epsilon$.
We again modify the contour to completely include the relevant domain
\begin{equation*}
    \int_0^{\infty}d\omega\,\omega^{\beta-1}\oint_{\cC(t+i\varepsilon)} \frac{d\mu}{2\pi i}\frac{\cT(z,\mu)}{\mu-t-i\varepsilon}\Bigl\vert_{t=-z\omega^2}= \int_{\cC(\bR^-+i\varepsilon)} \frac{d\mu}{2\pi i}\cT(z,\mu) \int_0^{\infty}d\omega\frac{\omega^{\beta-1}}{\mu+z\omega^2-i\varepsilon}
\end{equation*}
and we perform the Mellin integral of the free-propagator, obtaining
\begin{equation*}
    \int_0^{\infty}d\omega\,\omega^{\beta-1}\cT(s,t)\Bigl\vert_{\substack{s=\omega^2\\[0.05pt] t=-z\omega^2}} = \int_{\cC(\bR^-+i\varepsilon)} \frac{d\mu}{2\pi i}\cT(z,\mu)\Bigl(\frac{\pi z^{-\b/2}(-i\varepsilon+\mu)^{\b/2-1}}{2\sin(\pi\b/2)}\Bigr)\,.
\end{equation*}
where, similarly as before, we defined
\begin{equation*}
    \cT(z,\mu)=\cT(s,t)\vert_{\substack{s=s(z,\mu)\\[0.05pt] t=\mu\,\,\,\,\,\,\,\,\,}}\,.
\end{equation*}
Collecting all the contributions we get
\begin{equation}
\label{eq correlator t as contour integral}
    g(\beta,z) = \frac{\pi z^{2}}{2^{\beta+3}\sin(\pi\b/2)}\int_{\cC(\bR^-+i\varepsilon)} \frac{d\mu}{2\pi i}\, (\mu-i\varepsilon)^{\b/2-1}\,\cT(z,\mu)\,,
\end{equation}
where the regulator shifts the discontinuity from the $\mu^{\beta/2-1}$ slightly upwards. Notice that now the contour is taken along the  slightly shifted negative real line counterclockwise around $\mu=\bR^-+i\varepsilon$.

As anticipated in the previous section, we leave the regularization implicit in the following, with the understanding that the contour is always taken around the discontinuity and the poles and discontinuity coming from the amplitudes lie in the complementary region.

We can now apply the contour trick discussed to translate the Froissart-Gribov expansions \eqref{eq Partial wave expansion s-channel} and \eqref{eq Partial wave expansion t-channel} onto the celestial sphere. Notice that, since the partial waves depend only on $z$, the Mellin transform acts solely on the partial amplitudes $a_J$, and we can therefore apply the contour trick directly to them.

An important caveat, however, is that for this procedure to be well defined one must exchange the sum over spins arising from the partial wave expansion with the contour integral. As discussed in section \ref{sec: Regge limit}, this step is highly non-trivial. In particular, the $s$-channel Froissart-Gribov expansion generally converges for $z_s$ in the Lehmann-Martin ellipse, see \cite{Collins:1977jy} for further details. By contrast, for the $t$-channel expansion one must rely on different techniques, such as the Sommerfeld-Watson transform. This analysis will naturally lead us to the celestial Regge limit discussed in the next section.

We therefore consider first the Froissart-Gribov $s$-channel expansion. The discussion proceeds essentially as in \eqref{eq s-channel contour mellin transform}, with the appropriate modifications arising from the different kinematics. Consequently, the Froissart-Gribov expansion \eqref{eq Partial wave expansion s-channel} leads to the celestial expansion 
\begin{equation*}
    f_{\Delta_i}(z,\zb)= \delta(z-\zb)(1-z)^{\frac{\Delta_{12}-\Delta_{34}}{2}}z^{\beta/2+2}\frac{2^{-2-\beta}\pi}{\sin(\pi\beta/2)}\sum_{J=0}^{\infty} \tilde a_J(\beta) P_J(1-2z)\,.
\end{equation*}
This expression provides a celestial Froissart-Gribov expansion for the correlator. When comparing with the other known expansions, \ie conformal partial wave \cite{Nandan:2019jas, Atanasov:2021cje} and relativistic partial wave expansions \cite{Law:2020xcf}, this relation gives a precise bulk origin to the celestial data. In other terms, the complete knowledge of the partial amplitudes completely determine the celestial four-point functions on the celestial sphere.

Notice that all the dependence in $\mu$ is contained in the partial amplitudes. Therefore, the expression can be rewritten compactly with 
\begin{equation*}
\tilde a_J(\beta)=\int_{\cC^{-1}(\bR^++i\varepsilon)} \frac{d\mu}{2\pi i}(-\mu+i\varepsilon)^{\beta/2-1} a_J(s)\Bigl\vert_{s=\mu}\,,
\end{equation*}
where $\tilde a_J(\beta)$ are just the contour integral evaluated partial amplitudes. An analogous discussion leads to similar expression also for the $s$-channel kinematics; see section \ref{sec: Introduction and summary}.

\subsection{Mellin transform of the Regge amplitude}
Given the expression for the leading contribution to the amplitude in the Regge limit, \eqref{eq Regge limit of T}, we can perform the Mellin transform to obtain the celestial Regge limit. This procedure is reminiscent of the early derivations of Regge theory in CFT \cite{Costa:2012cb}. Starting from correlation functions in { position} space, one transforms to Mellin space, where the correlators take a form resembling scattering amplitudes. In this representation, poles in the conformal dimensions correspond to the exchange of operators.

In our case, we start from the amplitude in momentum space and perform the Mellin transform to obtain a correlation function on the celestial sphere. In order to do so we apply the Mellin transform using the alternative definition \eqref{eq correlator t as contour integral} to \eqref{eq Regge limit of T}. The reduced correlator in the Regge limit is then given by 

\begin{equation*}
    g^{\pm}(\beta,z) = \frac{\pi z^{2}}{2^{\beta+3}\sin(\pi\b/2)}\int_{\cC(\bR^-+i\varepsilon)} \frac{d\mu}{2\pi i}(\mu-i\varepsilon)^{\b/2-1} \Bigl(-\pi\,\rho^{\pm}(\mu)(-1)^{j(\mu)}\frac{z^{-j(\mu)}}{2^{1-2j(\mu)}} \frac{(1\pm (-1)^{{j(\mu)}})}{\sin(\pi j(\mu ))}\Bigr)\,,
\end{equation*} 
where we explicitly reintroduced the regulator for clarity. We then rewrite the two contributions in a clearer way by collecting in terms of sine and cosine
\begin{align*}
    \frac{1+(-1)^{j(\mu)}}{\sin(\pi j(\mu))} & =\frac{e^{i\pi j(\mu)/2}}{\sin(\pi j(\mu)/2)}\,,\\
    \frac{1-(-1)^{j(\mu)}}{\sin(\pi j(\mu))} & =\frac{-i \,e^{i\pi j(\mu)/2}}{\cos(\pi j(\mu)/2)}\,.
\end{align*}

The reduced celestial correlator \eqref{eq reduced correlator f of the t-kinematics} can be put in a suggestive form by writing it in terms of the norm of the cross-ratio $z=e^{i\theta}\zeta$. The overall delta function is then
\begin{equation*}
    \delta(z-\zb)= \delta(\zeta(e^{i\theta}-e^{-i\theta}))=\delta(2i\zeta\sin(\theta))\,.
\end{equation*}
Assuming $\zeta>0$, the delta is evaluated for $\theta=\{0,\pi,2\pi,..\}$. We can restrict it to the fundamental domain $\theta\in[0,2\pi)$. Moreover, note that for $\theta=\pi $, $z=-\zeta\implies z<0$, which is not possible considered that we take $\Re(z)\in (0,1)$. It follows that
\begin{equation*}
    \delta(z-\zb) = \frac{1}{2\zeta}\delta(\theta)\,.
\end{equation*}
Putting it all together the Regge limit of the 4-point amplitude in Mellin space can be written as
\begin{equation}
\label{eq celestial regge 4point}
    f^{\pm}_{\Delta_i}{(z,\zb)} \approx \delta(\theta)\,2\pi i\int_{\cC(\bR^-+i\varepsilon)} d\mu\,e^{i\pi j(\mu)} \alpha^{\pm}(\mu)\,\zeta^{1-j(\mu)}\,,
\end{equation}
where the contour goes around the negative real axis and $\alpha^{\pm}(\mu)$ are given by
\begin{align*}
    \alpha^{+}(\mu) &= e^{i\pi j(\mu)/2}\frac{\,2^{2 j(\mu)-\beta-7}\rho^+(\mu)\,{(-i\epsilon +\mu)}^{\beta/2-1}}{\pi\sin(\pi j(\mu)/2) \sin(\pi\beta/2)}\,,\\[0.3cm]
    \alpha^{-}(\mu) &= e^{i\pi j(\mu)/2}\frac{\,2^{2 j(\mu)-\beta-7}\,\rho^-(\mu)\,(-i\epsilon +\mu)^{\beta/2-1}}{i \pi\cos(\pi j(\mu)/2) \sin(\pi\beta/2)}\,.
\end{align*}
This expression defines the leading Regge contribution on the celestial sphere. Interestingly, as we will review in the next section, \eqref{eq celestial regge 4point} closely resembles the leading contribution in the Regge limit of a standard CFT for the special configuration $z=\bar z$, up to the different contour prescription adopted.

\begin{figure}
    \centering
    \includegraphics[width=0.45\linewidth]{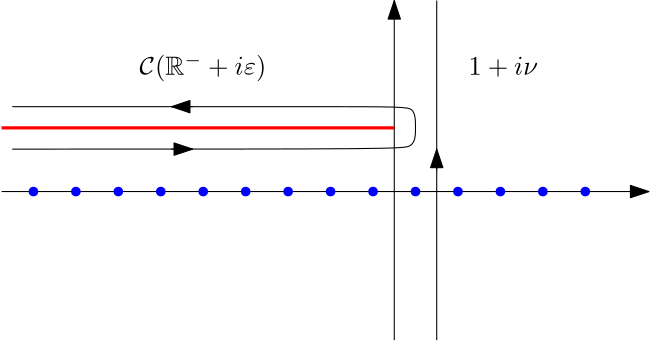}
    \caption{Contour modification from the discontinuity to the principal series. In blue the poles in $\alpha^\pm(\mu)$.}
    \label{fig: Regge-contour-trick}
\end{figure}

Motivated by this analogy, we now consider an alternative deformation of the contour that facilitates comparison with the standard CFT result. In particular, we focus on the contour associated with the principal series, deforming $\mathcal{C}(\mathbb{R}^-+i\varepsilon)$ into a line parallel to the imaginary axis, \ie $\mu = 1 + i\nu$ with $\nu\in\mathbb{R}$, as illustrated in figure \ref{fig: Regge-contour-trick}.

In order to actually perform such contour modification, we must require some assumptions on the the analytic structure of the $\alpha^{\pm}(\mu)$. We assume that $\rho^\pm(\mu)$ are holomorphic \footnote{at least until $\Re(\mu)\leq1$} and exponentially bounded;  see appendix \ref{app: regge pole}. Under this working hypothesis the only relevant contributions come from the poles of the sine and cosine at integer values of $j(\mu)$. Thus, by contour modification of \eqref{eq celestial regge 4point}, we get

\begin{equation}
\label{eq regge contour mod}
    f^{\pm}_{\Delta_i}{(z,\zb)} \approx \delta(\theta)\Bigl(\,4\pi^2 \sum_{\mu_l<1}\,e^{i\pi j(\mu_l)}\Res[ \alpha^{\pm}(\mu),\mu_l]\,\zeta^{1-j(\mu_l)} - \,2\pi \int_{-\infty}^{+\infty} d\nu\,e^{i\pi j(\nu)} \alpha^{\pm}(\nu)\,\zeta^{1-j(\nu)} \Bigr)\,,
\end{equation}
where $\mu_l$ are the corresponding value for the pole at integer $j(\mu_l)=l$. So far \eqref{eq regge contour mod} is rather general. We are interested in looking at which would be the leading contribution between the two factors in the Regge limit of $\zeta\ll1$. For that, we consider, as a working hypothesis, the spins to be arranged in linear trajectories that we parametrize as 
\begin{equation*}
    j(\mu)= j_0+ j_1\mu\,,
\end{equation*}
for some $j_0,j_1\in\bR$. This is for example true in string theory, as it can be seen by the Virasoro-Shapiro, but also in many phenomenological processes\footnote{See chapter 5 of \cite{Collins:1977jy} for examples}. With this parametrization, notice that the location of the poles is given by
\begin{equation*}
    j(\mu_l)=l\implies \mu_l = \frac{-j_0+l}{j_1}\qquad l\in\bZ\,,
\end{equation*}
where, for even signatures ($+$) $l$ is even and for odd ones ($-$) $l$ is odd. Notice that $\mu_l<1$  implies $l<j_0+j_1$.


As we will review in the next section, the leading contribution to the Regge limit in a standard CFT can be identified with the second term of \eqref{eq regge contour mod}. We therefore study, in the space of the parameters $j_0$ and $j_1$, which are the conditions under which we can actually disregard the first term. 

We begin by noticing that since the $\mu_l$ can grow in the negative real plane, in order for this contributions to be subleading in this limit, we must require that $j_1>0$ so that
\begin{equation}
\label{eq: poles contributio}
    \zeta^{1-j(\mu_l)}=e^{-(1-j_0-j_1 \mu_l)|\log(\zeta)|}\,,
\end{equation}
when evaluated at large negative $\mu_l$ converges for $\zeta\ll 1$. 

The second constraint thus comes by requiring that the second term grows as we lower $\zeta$. By a similar analysis, when we evaluate for $j(\mu)=j_0+j_1(1+i\nu)$ we have
\begin{equation}
\label{eq: regge contr}
    \zeta^{1-j(\mu)}=e^{-(1-j_0-j_1 -i\nu)|\log(\zeta)|}\,,
\end{equation}
which, beside an oscillatory term the contribution grows if
\begin{equation*}
    1-j_0-j_1<0\,.
\end{equation*}
We finally notice that \eqref{eq: regge contr} is leading by construction over \eqref{eq: poles contributio} as the requirement 
\begin{equation*}
    -1+j_0+j_1> -1+j_0+j_1\mu_l
\end{equation*}
 is just the condition $\mu_l<1$.
We thus see that the leading Regge term is given simply by

\begin{equation}
\label{eq Regge in the principal series}
    f^{\pm}_{\Delta_i}{(z,\zb)} \approx- \delta(\theta)\,2\pi \int_{-\infty}^{+\infty} d\nu\,e^{i\pi j(\nu)} \alpha^{\pm}(\nu)\,\zeta^{1-j(\nu)} \,.
\end{equation}
As we will see in the next section, this limit is closely related to that of the Regge limit of CFTs and will lead us to the identification between bulk quantities, such as the $\rho^{\pm}(\nu)$ and the conformal data. 

There are other natural approaches one could pursue from \eqref{eq celestial regge 4point}, that we briefly discuss. The first is to further deform the contour onto the branch cut and extract the corresponding discontinuity. This procedure, however, essentially reduces to the standard Mellin transform and the resulting integral cannot be computed without the explicit knowledge of the bulk data, such as the functions $\rho^\pm(\mu)$.

The second approach is to exploit the analytic structure of the amplitude and attempt to close the contour collecting all the residues of the poles and sum them. However, this strategy is not viable, as already indicated by the Regge behavior $\cA \sim s^{j(t)}$, since the contribution from the arcs at infinity does not vanish and therefore cannot be neglected.

\section{Partial amplitudes and OPE coefficients}
\label{sec:CFT Regge}
In this section we will review the standard Regge limit of CFTs, we will apply it to the celestial sphere and compare with the Regge limit of the celestial amplitude.

\subsection{Brief review of the Regge limit in CFT}

As we discussed in section \ref{sec: preliminaries}, the symmetry group that specifies the celestial CFT is that of a standard CFT plus additional symmetries, \ie the bulk translation invariance, which further constraint $f_{\Delta_i}(z,\zb)$. Hence, we may expect that the discussion of the Regge limit in celestial CFT should correspond to a particular sub-sector of that of a standard CFT having $SL(2,\bC)$ as defining symmetries.
Therefore, we first briefly recall the analysis of the Regge limit for CFTs \cite{Cornalba:2007fs, Costa:2012cb, Kravchuk:2018htv} showing how this can be extended to the celestial correlator \eqref{celestial_corr}, with some salient differences that we will note on the way. 

Therefore, we may begin from the four-point function of scalars, which can be generically expanded in conformal blocks.
\begin{equation*}
    \langle \phi_{h_1,\bar h_1}(z_1,\zb_1)\phi_{h_2,\bar h_2}(z_2,\zb_2)\phi_{h_3,\bar h_3}(z_3,\zb_3)\phi_{h_4,\bar h_4}(z_4,\zb_4)\rangle= K(z_i,\zb_i,h_i,\bar h_i)\sum_{\Delta,J} c^s_{\Delta,J} G_{\Delta,J}(z,\zb)\,,
\end{equation*}
by using the $SL(2,\bC)$ symmetry. Here, for simplicity, we define the $s$-channel OPE data by $c^s_{\Delta,J}=\lambda_{12\cO_{\Delta,J}}\lambda_{34\cO_{\Delta,J}}$, the functions $ G_{\Delta,J}(z,\zb)$ are the conformal blocks, see appendix \ref{app: sl2 CFT} for further details. In Euclidean signature we have the reality condition $\zb=z^*$, and this expansion is convergent for $|z|<1$. Indeed, for small values of the cross-ratios, the conformal blocks have a leading contribution given by
\begin{equation*}
    G_{\Delta,J}(z,\zb) \approx z^{\frac{\Delta-J}{2}}\zb^{\frac{\Delta+J}{2}}\,,
\end{equation*}
for $0\ll z\ll \zb \ll 1$.
In the following, we will be interested in a similar limit, but one that is crucially defined in a different setup, \ie in Lorentzian signature. After suitable manipulations, this will lead to the conformal Regge limit.

Following the standard discussion, and in order to make the connection with the celestial setup more transparent, we begin by briefly reviewing the analytic continuation of the partial-wave expansion. To this end, we consider either a standard CFT configuration, where the exchange of the identity operator is not treated explicitly, or a celestial CFT setup, in which the structure of the reduced correlator is fixed solely by imposing $SL(2,\mathbb{C})$ invariance. In either case, the conformally invariant function can be written as
\begin{equation}
\label{eq CPW expansion}
   f_{\Delta_i}(z,\zb)= \sum_{J=0}^{\infty}\int_{1-i\infty}^{1+i\infty}\frac{d\Delta}{2\pi i}c(\Delta,J) F_{\Delta,J}(z,\zb)
\end{equation}
where $\Delta=1+i\nu$, $c(\Delta,J)$ encodes the OPE data through deforming the contour and the $F_{\Delta,J}(z,\zb)$ are the combination of conformal blocks and shadows of the $SL(2,\bC)$, see appendix \ref{app: sl2 CFT} for more details. 
This expression typically arises from the so-called harmonic expansion. When evaluated at the poles in $\Delta$ of $c(\Delta,J)$, it reproduces the standard conformal block expansion. In conventional CFTs, the appearance of complex conformal dimensions is usually regarded as nonphysical and this representation is mainly used to expand the reduced correlator in a single-valued, orthogonal eigenbasis of the Casimir operators of the conformal group.

In celestial CFT, however, this basis has been argued to be physical, with exchanged operators belonging to the principal series $\Delta = 1 + i\mathbb{R}$ \cite{Pasterski:2017kqt}.
By combining with \eqref{reduced amplitude} and exploiting the orthogonality of the conformal waves, one can invert the expression, as argued in \cite{Nandan:2019jas}, and obtain the celestial CFT data $c(\Delta,J)$.

So, starting from \eqref{eq CPW expansion}, we aim to study the Regge limit in order to connect with the discussion of the previous section. Recall, however, that the Regge limit corresponds to an intrinsically Lorentzian causal configuration. To describe it, we must first Wick rotate to Lorentzian signature, \ie we take $z,\bar z \in \mathbb{R}$ as independent variables, obtained via analytic continuation of the Euclidean cross-ratios.

\begin{figure}
    \centering
    \includegraphics[width=0.4\linewidth]{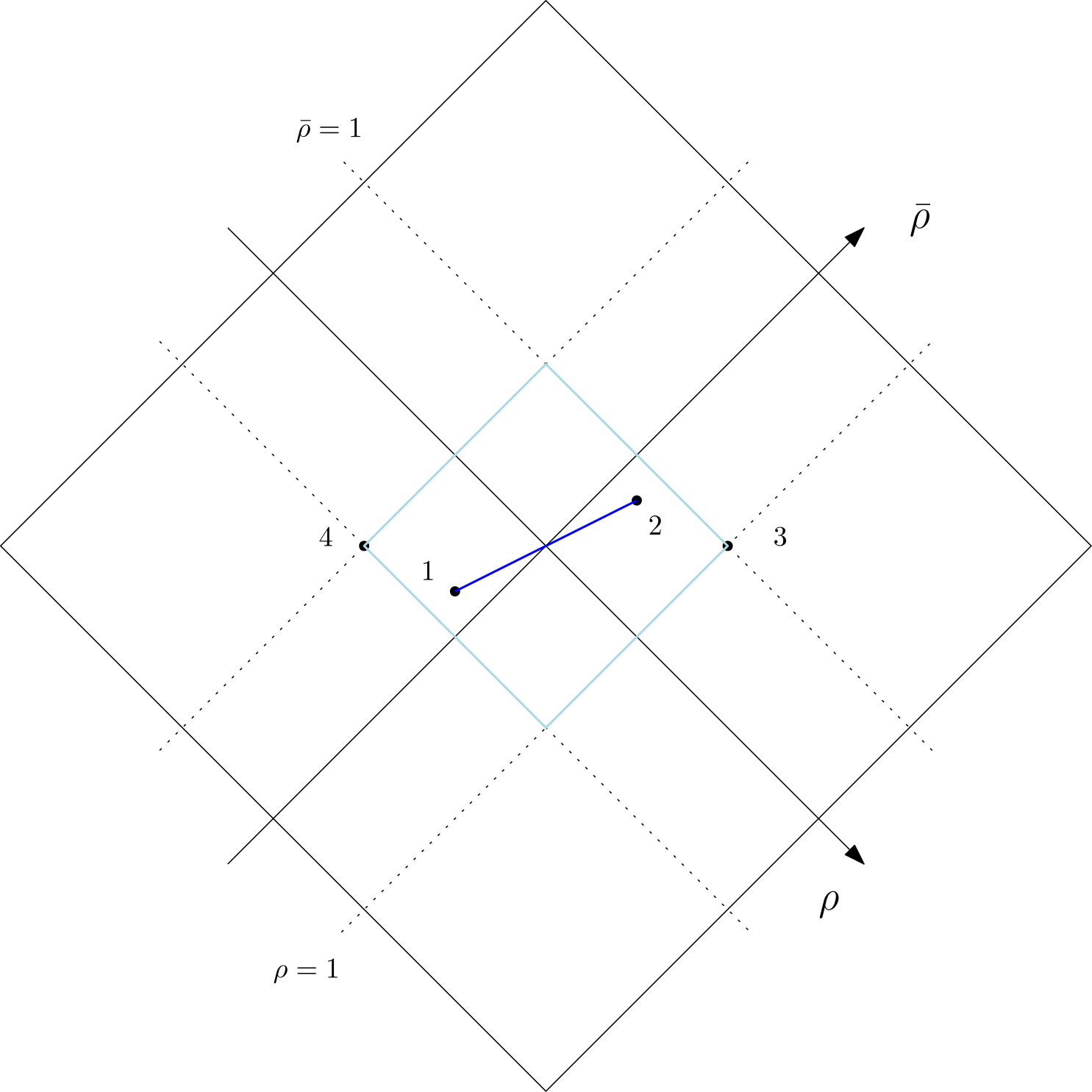}
    \caption{Initial configuration where all operators are spacelike separated and $1$ and $2$ are inside the causal diamond.}
    \label{fig:spacelike-config}
\end{figure}

In order to better grasp the analytic properties, it is best to move to the radial, or lightcone, coordinates $\rho,\bar\rho$, \cite{Hogervorst:2013sma, Caron-Huot:2017vep,Kravchuk:2018htv}. In particular, we consider the standard configuration where we take all the operators to be spacelike 
\begin{align*}
    x_1 &= (-\rho,-\bar\rho)\,,\\
    x_2 &= (\rho,\bar\rho)\,,\\
    x_3 &= (1,1)\,,\\
    x_4 &= (-1,-1)\,,
\end{align*}
as depicted in figure \ref{fig:spacelike-config}\footnote{We stress that this is not the diagram  of the standard celestial discussion. This is the conformal compactification of the $2d$ CFT and not of the $4d$ theory on the bulk.}.
In terms of the cross-ratio, we have that

\begin{equation*}
   \zb= \frac{4\bar\rho}{(1+\bar\rho)^2}\,,\qquad \,1-\zb=\frac{(1-\bar \rho)^2}{(1+\bar \rho)^2} \,,\qquad \bar\rho= \frac{1-\sqrt{1-\bar z}}{1+\sqrt{1-\bar z}}\,\,,
\end{equation*}
 and similarly for $z$ and $\rho$. 
In terms of these coordinates, the OPE converges for $\rho$ and $\bar{\rho}$ within the unit disc. From these expression, it will be clearer, later, how monodromies on $z$ or $\zb$, relates different configurations. For example, we see that taking $(1-\zb)\rightarrow e^{2\pi i}(1-\zb)$ produces $\bar\rho\rightarrow 1/\bar\rho$.

The procedure to obtain the Regge limit in CFT follows closely the discussion presented in the previous section. In particular, as argued in \cite{Costa:2012cb}, the analogy with the amplitude discussion is also more evident when one performs the comparison in Mellin space. Therefore, we first separate the contributions in definite signature and then we proceed with the Sommerfeld-Watson transformation. Thus, by trading the sums for a complex integral we obtain
\begin{equation*}
   f^{\pm}_{\Delta_i}(z,\zb)= \int_{1-i\infty}^{1+i\infty}\frac{d\Delta}{2\pi i}\int_{\cC_J} \frac{dJ}{2i \sin(\pi J)} c^{\pm}(\Delta,J) \Bigl(G_{\Delta,J}(z,\zb)\pm (-1)^J G_{\Delta,J}(z,\zb)\Bigr)
\end{equation*}
where the $c^{+}(\Delta,J)$ ($c^{-}(\Delta,J)$) refers to the even (odd) spin contributions when $J\in \bZ$, $\Delta=1+i\nu$ and the contour $\cC_J$ goes around the positive $\Re(J)$ axis. By assuming, just as in the bulk discussion, the presence of a leading isolated pole
\begin{equation*}
    c^{\pm}(\nu,J)\sim \frac{\sigma^{\pm}(\nu)}{J-j(\nu)}\,,
\end{equation*}
we can modify the contour to pick the residue at $J=j(\nu)$. Assuming that the contribution from the line integral is subleading, \cite{Cornalba:2007fs, Costa:2012cb, Kravchuk:2018htv}, we obtain 
\begin{equation*}
   f^{\pm}_{\Delta_i}(z,\zb)\approx -\int_{-\infty}^{\infty}\frac{d\nu}{2\pi} \frac{\pi\,\sigma^{\pm}(\nu)}{\sin(\pi j(\nu))}\Bigl(G_{1+i\nu,j(\nu)}(z,\zb)\pm (-1)^{j(\nu)} G_{1+i\nu,j(\nu)}(z,\zb)\Bigr)\,,
\end{equation*}
where the overall minus comes from the clock-wise contour around the pole. 

\begin{figure}
    \centering
    \includegraphics[width=0.4\linewidth]{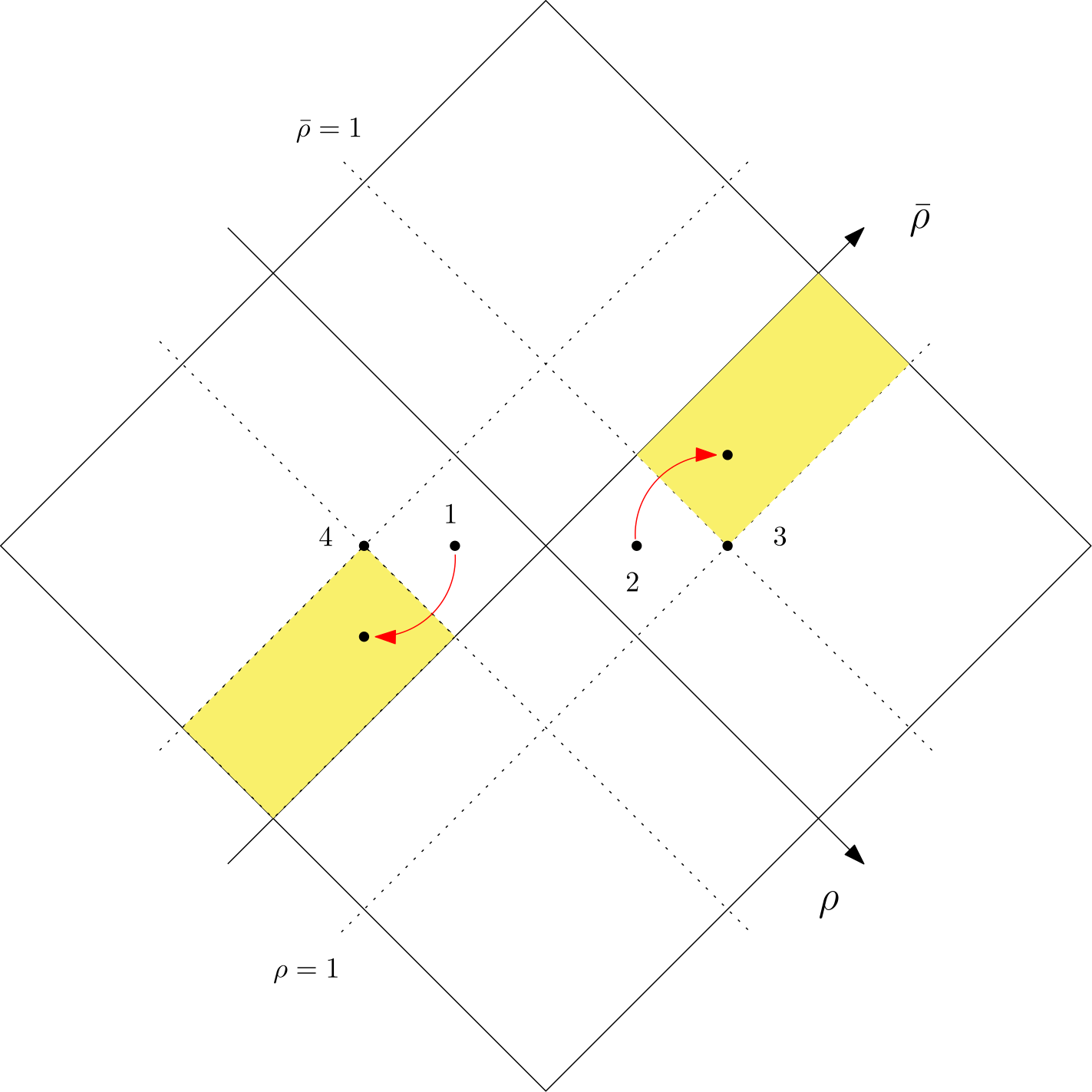}
    \caption{Lorentzian configuration where $1$ and $2$ are sent to the lightcones of $4$ and $3$ respectively. In red, the effect of taking the monodromy $(1-\zb)\rightarrow e^{2\pi i }(1-\zb)$ or, equivalently, $\bar\rho\rightarrow1/\bar \rho $}
    \label{fig:analytic-continuation}
\end{figure}

\begin{figure}
    \centering
    \includegraphics[width=0.4\linewidth]{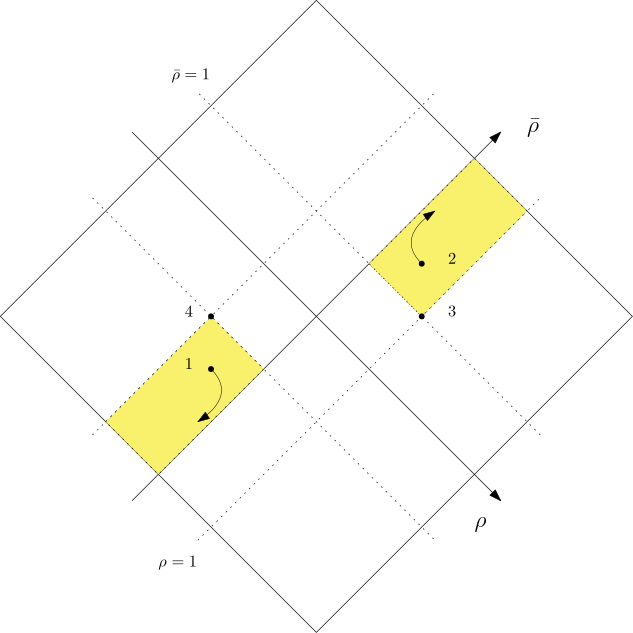}
    \caption{Causal configuration specified by $0<\rho <1<\bar\rho<\infty$. In black the Regge limit associated to small cross-ratios $z,\,\zb$. }
    \label{fig: Regge-limit}
\end{figure}

We consider now the analytic continuation.
Since at the end of the process we want to meet the celestial sphere where the correlator is evaluated on the real line, \ie $z=\zb$, we can analogously take the analytic continuation in $z$ or $\zb$. After Wick rotating to Lorentzian signature, we consider, as typical in this discussion, the counterclockwise continuation around $\zb=1$ thus obtaining
\begin{equation}
\label{eq Regge in CFT}
    f^{\pm}_{\Delta_i}(z,\zb)\approx -\int_{-\infty}^{\infty}{d\nu}\,\sigma^{\pm}(\nu) \frac{1\pm (-1)^{j(\nu)}}{2\sin(\pi j(\nu))}G_{1+i\nu,j(\nu)}^{\circlearrowleft}(z,\zb)\,.
\end{equation}
This configuration corresponds to placing two operator insertions on the light cone of the others while keeping the two pairs spacelike separated, as shown in figure~\ref{fig:analytic-continuation}. Operator $1$ lies in the past of $4$, and operator $2$ lies in the future of $3$. We denote this causal ordering with the notation $4>1$ and $2>3$, where $>$ indicates causal separation, \cite{Kravchuk:2018htv}.
After the analytic continuation, we can now take the limit for $z,\zb\rightarrow 0$ at fixed ratio, figure \ref{fig: Regge-limit}.

The explicit evaluation of the monodromy, performed in appendix \ref{app sub: Monodromies of the conformal blocks}, leads to
\begin{equation}
\label{eq regge in cft line int}
    f^{\pm}_{\Delta_i}(z,\zb)\approx - \,2\pi \int_{-\infty}^{+\infty} d\nu\,e^{i\pi j(\nu)} \gamma^{\pm}(\nu)\,\zb^{1-j(\nu)} \Bigl(\frac{\zb}{z}\Bigr)^{\frac{j(\nu)-\Delta(\nu)}{2}}\,,
\end{equation}
where, in order to keep a streamlined expression, we keep some contributions in terms of $\Delta(\nu)=1+i\nu$. The coefficients $\gamma^{\pm}(\nu)$ encodes the conformal data and are defined by
\begin{align}
    \gamma^{+}(\nu) &= \frac{e^{-i\pi j(\nu)/2}\,\sigma^{+}(\nu)}{4\sin(\pi j(\nu)/2)}\frac{ e^{\pi i\frac{\Delta_{12}-\Delta_{34}}{2}}}{i \pi^2\, K_{\Delta(\nu),j(\nu)}}\,,\\
    \gamma^{-}(\nu) &= \frac{e^{-i\pi j(\nu)/2}\,\sigma^{-}(\nu)}{4\, i\cos(\pi j(\nu)/2)}\frac{e^{\pi i\frac{\Delta_{12}-\Delta_{34}}{2}}}{i\pi^2\, K_{\Delta(\nu),j(\nu)}}\,,
\end{align}
where $K_{\Delta,J}$ is that defined in \eqref{eq K gamma combination}.

\subsection{Connecting the CFT and CCFT Regge limits}
In this last subsection, we connect the Regge limit in ordinary CFT with the Regge limit in celestial CFT. Already from \eqref{eq regge in cft line int} one can observe a close similarity with the expression obtained from the Mellin transform of the bulk Regge limit, \ie \eqref{eq Regge in the principal series}. One of the main differences with the standard CFT case is the additional constraint imposed by the bulk momentum conservation. 
Nevertheless, since CCFT enjoys an enhanced symmetry structure, it can be regarded as a constrained instance of a CFT. In particular, the conformal invariant function $f_{\Delta_i}(z,\zb)$ appearing in \eqref{celestial_corr} is completely fixed by $SL(2,\bC)$ symmetry, while the full celestial correlator is further restricted by the $\bR^{1,3}$ invariance. In both frameworks, the correlator depends on the same cross-ratio. Consequently, analytic continuations in cross-ratio space, such as those defining the Regge limit in CFT, must be mirrored in celestial CFT.
This observation motivates us to carry out the same analytic manipulations in both settings in order to make their relation manifest.

Notice, however, that this construction is motivated purely by an educated guess. Indeed, adopting a bottom-up approach and lacking an independent definition of a celestial (distributional-like) CFT, the prescription one chooses in order to match the physical data is, to some extent, arbitrary. Nevertheless, we remark that \eqref{eq celestial regge 4point} still defines a proper Mellin transform of the Regge limit. Thus, although different prescriptions may lead to different overall multiplicative factors, \eg phases, the similarities between the Regge behavior and hence a final relation between the (residue) OPE data and the partial amplitudes is expected to hold in general.

We can now apply all this discussion to our case of the celestial correlator with the generic expression \eqref{reduced amplitude} 
\begin{equation*}
    f_{\Delta_i}(z,\zb) =\delta(z-\zb)(1-z)^{\frac{\Delta_{12}-\Delta_{34}}{2}}g(z)\,.
\end{equation*}

To relate the two Regge limits, we apply to \eqref{eq Regge in the principal series} the same analytic continuation in cross-ratio space that leads to \eqref{eq regge in cft line int} in the ordinary CFT setting. Since both expressions depend on the same cross-ratio, consistency requires that the analytic continuation be implemented in an identical manner.

The first obstacle of applying these arguments directly is the non-analytic nature of the delta function. To address this issue, we adopt a prescription that allows us to define and take the monodromy in the presence of the delta function, as discussed in Appendix~\ref{app: Dealing with the distributions}.

Therefore, after the analytic continuation to Lorentzian we take the monodromy on each side
\begin{equation*}
    M_{\zb=1}[f_{\Delta_i}(z,\zb)] = \delta(z-\zb)M_{\zb=1}[(1-z)^{\frac{\Delta_{12}-\Delta_{34}}{2}}g(z)]\,.
\end{equation*}
Here, $g(z)$ is actually monodromy free because it is written from $\cT(s,t)$ as an expansion in Chebychev polynomials, which are monodromy free. 

We thus have
\begin{equation*}
    M_{\zb=1}[f_{\Delta_i}(z,\zb)] = \delta(z-\zb)M_{z=1}[(1-z)^{\frac{\Delta_{12}-\Delta_{34}}{2}}]g(z)\,.
\end{equation*}
which simply gives 
\begin{equation*}
    M_{\zb=1}[f_{\Delta_i}(z,\zb)] = \delta(z-\zb)e^{i\pi(\Delta_{12}-\Delta_{34})}(1-z)^{\frac{\Delta_{12}-\Delta_{34}}{2}}g(z)\,.
\end{equation*}
From this the limit of small cross-ratio can be taken.

\subsubsection{Going back to Euclidean}

However, one final step is required in order to meaningfully compare the two results. Since the two Regge limits are naturally defined in Lorentzian signature, obtaining a Euclidean result requires analytically continuing back and restoring $\bar z = z^\ast$.

\begin{figure}
    \centering
    \includegraphics[width=0.4\linewidth]{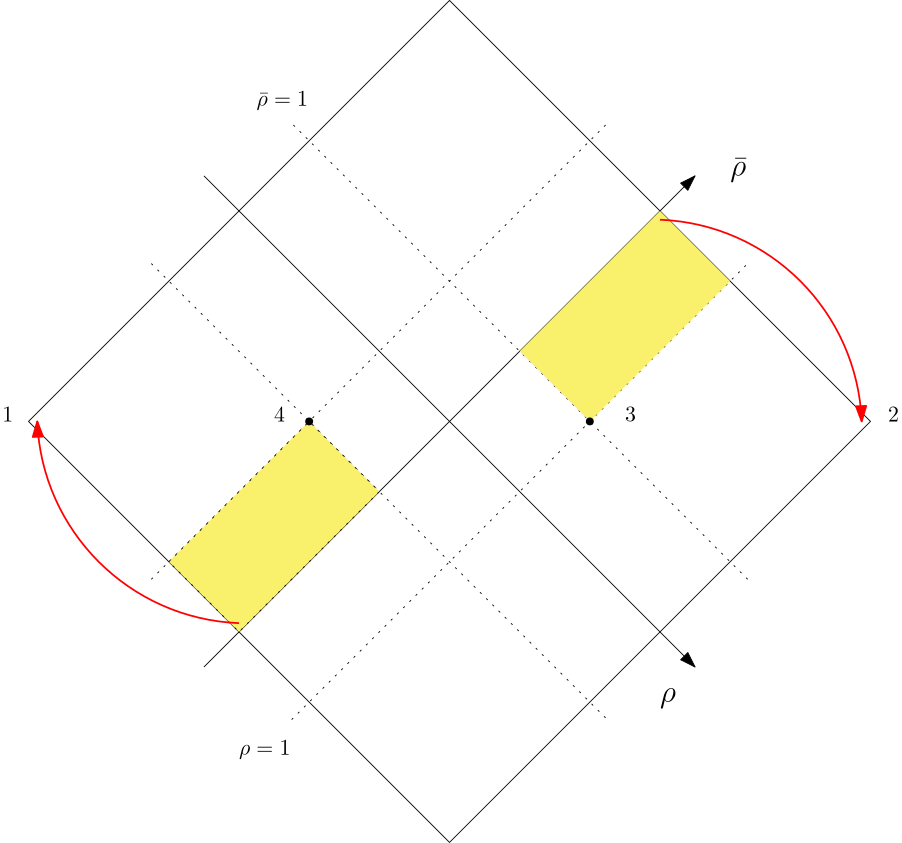}
    \caption{Effect of mapping back to spacelike configuration, $\rho\rightarrow1/\rho$, after the Regge limit}
    \label{fig: Second-analytic-continuation}
\end{figure}

To this end, we first consider an analytic continuation that maps the configuration to a causal setup in which all operators are spacelike separated. This is achieved by continuing around $z=1$, which, in terms of the radial coordinates $\rho,\bar\rho$, corresponds to the transformation $\rho \rightarrow 1/\rho$. Together with the small-cross-ratio limit, this maps the Regge configuration, where the two pairs of insertions are causally related, to a configuration in which all operators lie on the real line and are mutually spacelike, as illustrated in Figure~\ref{fig: Second-analytic-continuation}. 
This continuation sends operators $1$ and $2$ to $i^0$, \ie spacelike infinity, with one approaching from $\mathcal{I}^{-+}$ and the other from $\mathcal{I}^{+-}$.

From this configuration, the Euclidean setup can be straightforwardly recovered by a Wick rotation.

A second crucial outcome of this analysis is that it removes the ambiguity in how one analytically continues to Lorentzian time and in which of the two cross-ratios one performs the (first) analytic continuation followed by the Regge limit. In standard AdS/CFT, this choice is dictated by the physical setup under consideration, \eg to match a bulk analysis or a specific causal ordering. In our case, however, this continuation is merely an intermediate step.
Indeed, the final expression \eqref{eq regge in cft line int} cannot be naively Wick rotated back to Euclidean signature, as this would break holomorphicity. As a consequence, the final Euclidean (C)CFT result must be independent of whether one considers the intermediate ordering $4>1$ and $2>3$, or the opposite $1>4$ and $3>2$.

Notice that the same analytic continuation must also be implemented on the celestial CFT side. However, from the explicit form of the leading term in the celestial Regge limit, one can see that this final transformation does not modify the structure of \eqref{eq Regge in the principal series}. After this step, we can rotate back to the Euclidean configuration.

Accordingly, after returning to Euclidean signature, we write $z=\zeta\,e^{i\theta}$ and we restrict to the real line by setting $z=\zb$. We emphasize this feature by rewriting the expression in a distributional vest, exploiting the relation \eqref{eq distrib eq}, thus obtaining
\begin{equation*}
    \mathfrak{f}^{\pm}_{\Delta_i}(z,\zb)\approx - \,2\pi\delta(\theta)\int_{-\infty}^{+\infty} d\nu\,e^{i\pi j(\nu)} \gamma^{\pm}(\nu)\,\zeta^{1-j(\nu)} \,,
\end{equation*}
where $\mathfrak{f}^{\pm}_{\Delta_i}(z,\zb)= f^{\pm}_{\Delta_i}(z,\zb)\vert_{z=\zb}$ with $ f^{\pm}_{\Delta_i}(z,\zb)$ that of \eqref{eq regge in cft line int}. The expression is thus evaluated on the support of $\theta=0$, essentially leading us to a Euclidean CFT Regge limit. 

Summarizing, we started from the correlation functions of the celestial CFT and the $SL(2,\bC)$ CFT and we first rotated from Euclidean to Lorentzian signature. In the Lorentzian setup, we then performed the analytic continuation associated with $\bar \rho \rightarrow 1/\bar \rho$, namely $(1-\bar z)\rightarrow(1-\bar z)e^{2\pi i}$, and subsequently took the limit $\bar z\rightarrow 0$ while keeping the ratio $\bar z / z$ fixed.
Next, we considered a second continuation, $\rho \rightarrow 1/\rho$. This transformation did not modify the expression for the Regge limit of the celestial CFT; rather, it affected the standard Regge limit by mapping the causally ordered configuration to a spacetime configuration in which all operators are aligned, thereby producing a Dirac delta. At this stage, the configuration was still Lorentzian, so we finally rotate back to Euclidean signature. Since all operators are spacelike separated, this last rotation is straightforward.

We can now compare the expression with that obtained from the celestial CFT \eqref{eq Regge in the principal series}, finding a remarkable equivalence. This comparison allows for the identification between the conformal data obtained by a CFT discussion with the celestial data which are directly related to the Regge residues of the bulk partial amplitudes
\begin{equation*}
    \gamma^{+}(\nu)=e^{i\pi(\Delta_{12}-\Delta_{34})}\alpha^+(\nu)\,,\qquad \gamma^{-}(\nu)=e^{i\pi(\Delta_{12}-\Delta_{34})}\alpha^-(\nu)\,,
\end{equation*}
which leads to the relations
\begin{align*}
    \frac{-i\, e^{\pi i\frac{\Delta_{12}-\Delta_{34}}{2}}}{\pi^2 K_{\Delta(\nu),j(\nu)}}\frac{e^{-i\pi j(\nu)/2}\,\sigma^{+}(\nu)}{4\sin(\pi j(\nu)/2)} &= e^{i\pi j(\mu)/2}\frac{\,2^{2 j(\mu)-\beta-7}\rho^+(\mu)\,\mu^{\beta/2-1}\,e^{i\pi(\Delta_{12}-\Delta_{34})}}{\pi\sin(\pi\beta/2)\sin(\pi j(\mu)/2)}\,,\\[0.3cm]
    \frac{-i\,e^{\pi i\frac{\Delta_{12}-\Delta_{34}}{2}}}{\pi^2 K_{\Delta(\nu),j(\nu)}}\frac{e^{-i\pi j(\nu)/2}\,\sigma^{-}(\nu)}{4i\,\cos(\pi j(\nu)/2)} &= e^{i\pi j(\mu)/2}\frac{\,2^{2 j(\mu)-\beta-7}\,\rho^-(\mu)\,\mu^{\beta/2-1}\, e^{i\pi(\Delta_{12}-\Delta_{34})}}{\pi i\,\sin(\pi\beta/2)\cos(\pi j(\mu)/2)}\,,
\end{align*}
where the \rhs is evaluated in the principal series $\mu=1+i\nu$.
Simplifying we finally obtain

\begin{equation}
\label{eq residues of OPE = residues of partial ampltiude}
    \sigma^{\pm}(\nu) = i\pi\, e^{2\pi i j(\nu)}\frac{2^{2j(\nu)-\beta-5}\Delta(\nu)^{\beta/2-1}}{\sin(\pi\beta/2)}e^{\pi i\frac{\Delta_{12}-\Delta_{34}}{2}}K_{\Delta(\nu),j(\nu)}\rho^{\pm}(\nu)\,,
\end{equation}
where, given that $\mu=1+i\nu$ is the same as the expression for $\Delta(\nu)=1+i\nu$, we wrote it in the more suggestive way with $\Delta(\nu)=\mu(\nu)$.

Up to overall factors, which depend on the prescription adopted and on the two different procedures used to obtain the Regge limit, we find an equivalence between conformal data and bulk data. Moreover, $\rho^{\pm}(\nu)$ generically arise from quantum effects. As a result, this expression encodes dynamical information beyond tree-level contributions, providing further evidence that genuinely dynamical data are captured by the celestial CFT.
Since these are residues, we can reintroduce the poles in $J=j(\nu)$ on both sides and write the matching as
\begin{equation}
\label{eq OPE = partial ampltiude}
    c^{\pm}(\nu,J) \approx i\pi \,e^{2\pi i J}\frac{2^{2J-\beta-5}\Delta(\nu)^{\beta/2-1}}{\sin(\pi\beta/2)}e^{\pi i\frac{\Delta_{12}-\Delta_{34}}{2}}K_{\Delta(\nu),J}\,a_J^{\pm}(\nu)\,.
\end{equation}
with the caveat that the above is true up to possible terms that vanish at the poles. If we consider all the contributions, \ie all the subleading contributions coming from the line integrals, is not guaranteed that the equality holds. A full match requires the study of all analytic properties of these quantities.

\section{Discussion}
\label{sec:discussion}

Conformal partial-wave expansions and their associated OPE coefficients have been studied by several authors \cite{Atanasov:2021cje,Chang:2023ttm,Pacifico:2025emk,Surubaru:2025qhs,Fan:2023lky,Fan:2021pbp}. However, the relationship between bulk and boundary quantities remains not fully understood. In this context, our results in \eqref{eq residues of OPE = residues of partial ampltiude} or, equivalently \eqref{eq OPE = partial ampltiude}, establish a relation between independently defined bulk and boundary quantities in the Regge limit.

This equivalence follows crucially from the contour deformations employed to compute the Mellin transform, introduced in section \ref{sec:celestial Regge}. We expect that these methods may also prove useful in more general kinematic regimes.

For gravity, the Regge amplitude in the bulk is not the dominant contribution in this limit. It is the eikonal amplitude that dominates and some studies on celestial eikonal amplitudes have appeared in the literature \cite{Adamo:2024mqn,deGioia:2022fcn}. It would be interesting to find a relation between them and the Regge limit on the celestial sphere as we have defined in section \ref{sec:CFT Regge}.

The CFT operators that are associated to the leading contribution in the CFT Regge limit are the Lorentzian light-ray operators \cite{Kravchuk:2018htv,Kobayashi:2020kgb}. Some of these non-local operators can be written as integrals of local fields along a light-like direction, which calls to mind the half-shadow operators which have appeared before in the study of CCFTs \cite{Jorge-Diaz:2022dmy,Banerjee:2022hgc,Hu:2022syq,Sharma:2021gcz,Banerjee:2025oyu,Himwich:2025bza,Himwich:2025ekg}. We will address the relation between half-shadow operators and the celestial Regge limit in future work.

\subsection*{Acknowledgments}

This study was financed, in part, by the São Paulo Research Foundation (FAPESP), Brasil. Process Number 2024/12765-6.
\appendix

\section{Celestial four-point function}
\label{app: celestial 4-point function}
We briefly review how the celestial correlation functions are obtained from the bulk amplitude in the different kinematics. 

As for the standard CFT computations, we begin by setting up the correlator in the conformal frame
\begin{equation*}
    z_1=0\,\,\qquad z_2=z\,,\qquad z_3=1\,,\qquad z_4 = \Lambda\,,
\end{equation*}
with $\Lambda\rightarrow\infty$, and similarly for the the conjugate one.
In this setup, we can explicit the delta in the quantity: $z$, $\omega_1$, $\omega_2$ and $\omega_3$. In this respect, we have to rewrite the momentum conservation in these variables. We do that by taking into account the Jacobian
\begin{equation*}
   ( J )^{i}_{j}= \frac{\partial p_{tot}^{i-1}}{\partial x^j}\,,
\end{equation*}
with $i,j=1,...,4$ and $\vec x = (\omega_1,\omega_2,\omega_3,z)$ and $p_{tot}^\mu= \sum_{k=1}^4\,p_k^\mu$. Given that
\begin{equation*}
    \det J= 4i\zb(\zb-1)\eta_1\eta_2^2\eta_3\,\omega_2\,,
\end{equation*}
the delta can be rewritten as
\begin{equation*}
    \delta^{(4)}(p_{tot})= \frac{\delta(\omega_1-\frac{(\Lambda-1)(\zb-\bar\Lambda)\eta_4\omega_4}{\zb \eta_1})\delta(\omega_2-\frac{(\zb\Lambda+\bar\Lambda-(\zb+\Lambda)\bar\Lambda)\eta_4\omega_4}{\zb(\zb-1) \eta_2})\delta(\omega_3-\frac{(\bar\Lambda-\zb)\Lambda)\eta_4\omega_4}{(\zb-1) \eta_3})\delta(z-\frac{\zb\Lambda(\bar\Lambda-1)}{\zb\Lambda+\bar\Lambda-(\Lambda+\zb)\bar\Lambda})}{|4i\zb(\zb-1)\eta_1\eta_2^2\eta_3\,\omega_2|}\,,
\end{equation*}
where the denominator comes from the  Jacobian of the transformation. If we take $\Lambda,\bar\Lambda\gg1$ we can keep only the leading contribution, thus obtaining
\begin{equation*}
    \delta^{(4)}(p_{tot})= \frac{1}{|4i\zb(\zb-1)\eta_1\eta_2^2\eta_3\,\omega_2|}{\delta\Bigl(\omega_1+\frac{\Lambda\bar\Lambda\eta_4\omega_4}{\zb \eta_1}\Bigr)\delta\Bigl(\omega_2+\frac{\Lambda\bar\Lambda\eta_4\omega_4}{\zb(\zb-1) \eta_2}\Bigr)\delta\Bigl(\omega_3-\frac{\bar\Lambda\Lambda\eta_4\omega_4}{(\zb-1) \eta_3}\Bigr)\delta(z-\zb)}\,.
\end{equation*}
 
Notice that the requirement of each energy being positive, produce the constraints
\begin{equation*}
    \Theta\Bigl(-\frac{\eta_4\eta_1}{z}\Bigr)\Theta\Bigl(-\frac{\eta_4\eta_2}{z(z-1)}\Bigr)\Theta\Bigl(\frac{\eta_4\eta_3}{(z-1))}\Bigl)\,,
\end{equation*}
so, for example, $\eta_1=\eta_3=-1$ implies $z\in[0,1]$. 
The Mandelstam invariants are 
\begin{align}
    s_{12}=-(p_1+p_2)^2&=\frac{4\Lambda^2\bar\Lambda^2\eta_4^2\omega_4^2}{z-1}\,,\nonumber\\
    s_{13}=-(p_1+p_3)^2&=\frac{4\Lambda^2\bar\Lambda^2\eta_4^2\omega_4^2}{z(1-z)}\,,\nonumber\\
    s_{14}=-(p_1+p_4)^2&=\frac{4\Lambda^2\bar\Lambda^2\eta_4^2\omega_4^2}{z}\,.\nonumber
\end{align}
Keep in mind that here, as in the main text, we define $s$ to be always the center of mass energy, regardless of the kinematics. So in the $s$-kinematics $s=s_{12}$, but in the $t$-kinematics we call $s=s_{13}$. 

We can rewrite the integral
\begin{equation*}
    \int_0^\infty\prod_{i=1}^{4}d\omega_i\,\omega^{\Delta_i-1}\delta^{(4)}(p_{tot})\cT(s,t)
\end{equation*}
by explicitly integrating out the Dirac deltas
\begin{align*}
   \int_0^\infty &\frac{d\omega_4}{4\Lambda^2}(z-1)^{2-\Delta_2-\Delta_3}\,z^{2-\Delta_1-\Delta_2}\Lambda^{2(-3+\Delta_1+\Delta_2+\Delta_3)}\,\eta_1^{1-\Delta_1}\eta_2^{1-\Delta_2}\eta_3^{1-\Delta_3}\eta_4^{-3+\Delta_1+\Delta_2+\Delta_3}\,\\ &\omega_4^{\Delta_1+\Delta_2+\Delta_3+\Delta_4-5}\cT(\frac{4\Lambda^4\eta_4^2\omega_4^2}{z-1},\frac{4\Lambda^4\eta_4^2\omega_4^2}{z(1-z)})\,,
\end{align*}
where we just took $\Lambda=\bar\Lambda$. Depending on the channel one obtains different outcomes. If we take the $s$-channel kinematics $-\eta_1=-\eta_2=\eta_3=\eta_4=1$, then
\begin{equation*}
    \omega=\frac{2\Lambda^2\omega_4}{\sqrt{z-1}}\,,
\end{equation*}
with $z>1$. We thus obtain
\begin{equation*}
  \prod_i\eta_i^{1-\Delta_i}\, z^{-\Delta_1-\Delta_2}\,\Lambda^{-2\Delta_4} \delta(z-\zb)(z-1)^{\frac{\Delta_{12}-\Delta_{34}}{2}}\,2^{-2-\beta}z^2\int_0^{\infty} d\omega \,\omega^{\beta-1}\cT(s,t)\Bigr\vert_{s=\omega^2,\, t=-\omega^2/z}
\end{equation*}
where, given the considered kinematics, the prefactor is $\prod_i\eta_i^{1-\Delta_i}=(-1)^{-\Delta_1-\Delta_2}$. If we normalize for the overall, we get the massless scalar celestial correlator
\begin{equation*}
    \cA^{(4)}(z_i,\zb_i;\Delta_i)= z^{-\Delta_1-\Delta_2}\,\Lambda^{-2\Delta_4}\delta(z-\zb)(z-1)^{\frac{\Delta_{12}-\Delta_{34}}{2}}\,2^{-2-\beta}z^2\int_0^{\infty} d\omega \,\omega^{\beta-1}\cT(s,t)\Bigr\vert_{s=\omega^2,\, t=-\omega^2/z}\,,
\end{equation*}
with $z\in[1,\infty)$. As we are about to see the prefactor $z^{-\Delta_1-\Delta_2}\,\Lambda^{-2\Delta_4}$ is the very same of the conformal frame of the standard CFT.
For the different $t$-kinematics we can perform similar computations thus obtaining
\begin{equation*}
    \cA^{(4)}(z_i,\zb_i;\Delta_i)= z^{-\Delta_1-\Delta_2}\,\Lambda^{-2\Delta_4}\delta(z-\zb)(1-z)^{\frac{\Delta_{12}-\Delta_{34}}{2}}\,2^{-2-\beta}z^{\frac{\beta}{2}+2}\int_0^{\infty} d\omega \,\omega^{\beta-1}\cT(s,t)\Bigr\vert_{s=\omega^2,\, t=-z\omega^2}\,.
\end{equation*}
with now $z\in[0,1]$.

We want now to relate with the standard expression of a $2d$ CFT where
\begin{align*}
    \langle \phi_{h_1,\bar h_1}(z_1,\zb_1 )&\phi_{h_2,\bar h_2}(z_2,\zb_2 )\phi_{h_3,\bar h_3}(z_3,\zb_3) \phi_{h_4,\bar h_4}(z_4,\zb_4) \rangle=\\
    &=\frac{1}{z_{12}^{h_1+h_2}z_{34}^{h_3+h_4}}\Bigl(\frac{z_{14}}{z_{13}}\Bigr)^{h_{34}}\Bigl(\frac{z_{24}}{z_{14}}\Bigr)^{h_{12}} \frac{1}{\zb_{12}^{h_1+h_2}\zb_{34}^{h_3+h_4}}\Bigl(\frac{\zb_{14}}{\zb_{13}}\Bigr)^{\bar h_{34}}\Bigl(\frac{\zb_{24}}{\zb_{14}}\Bigr)^{\bar h_{12}}f(z,\zb)\,.
\end{align*}
If we evaluate the expression in the same conformal frame as above we get
\begin{align*}
    \langle \phi_{h_1,\bar h_1}(0,0)&\phi_{h_2,\bar h_2}(z, \zb )\phi_{h_3,\bar h_3}(1,1) \phi_{h_4,\bar h_4}(\Lambda,\bar\Lambda) \rangle=\\
    &=(-1)^{-\Delta_1-\Delta_2-\Delta_3-\Delta_4}\Lambda^{-J_4-\Delta_4}\bar\Lambda^{J_4-\Delta_4}\, z^{-\frac{J_1+J_2+\Delta_1+\Delta_2}{2}}\zb^{\frac{J_1+J_2-\Delta_1-\Delta_2}{2}}f(z,\zb)\,.
\end{align*}
Evaluating on the real line one obtains
\begin{equation*}
    \langle \phi_{h_1,\bar h_1}(0,0)\phi_{h_2,\bar h_2}(z, \zb )\phi_{h_3,\bar h_3}(1,1) \phi_{h_4,\bar h_4}(\Lambda,\bar\Lambda) \rangle\Bigr\vert_{z=\zb}=(-1)^{-\Delta_1-\Delta_2-\Delta_3-\Delta_4}\Lambda^{-2\Delta_4} z^{-\Delta_1-\Delta_2}f(z,z)\,.
\end{equation*}
We can therefore set the equation relating the CFT and CCFT result, thus leading to 
\begin{align*}
    (-1)^{-\Delta_1-\Delta_2-\Delta_3-\Delta_4}&\Lambda^{-2\Delta_4} z^{-\Delta_1-\Delta_2}f(z,z)=\\
    (-1)^{-\Delta_1-\Delta_2}\,z^{-\Delta_1-\Delta_2}\,&\Lambda^{-2\Delta_4} \delta(z-\zb)(z-1)^{\frac{\Delta_{12}-\Delta_{34}}{2}}\,2^{-2-\beta}z^2\int_0^{\infty} d\omega \,\omega^{\beta-1}\cT(s,t)\Bigr\vert_{s=\omega^2,\, t=-\omega^2/z}\,,
\end{align*}
where we have on the \lhs the reduced correlator of a $SL(2,\bC)$ CFT and on the \rhs with the additional $4d$ bulk momentum conservation which produces the delta.
Notice that under this perspective the Euclidean correlator can be regarded as with all incoming legs. If we normalize all the external operator by these (incoming) phases, we have a clear match between the two sides where
\begin{equation*}
    f(z,z) =(z-1)^{\frac{\Delta_{12}-\Delta_{34}}{2}}\,2^{-2-\beta}z^2\int_0^{\infty} d\omega \,\omega^{\beta-1}\cT(s,t)\Bigr\vert_{s=\omega^2,\, t=-\omega^2/z}\,.
\end{equation*}
We can perform the very same analysis now for the $t$-kinematics, \ie with the first and third particle incoming, thus obtaining a similar expression
\begin{equation*}
    f(z,z)=z^{-\Delta_1-\Delta_2}\,\Lambda^{-2\Delta_4} (1-z)^{\frac{\Delta_{12}-\Delta_{34}}{2}}\,2^{-2-\beta}z^{\frac{\beta}{2}+2}\int_0^{\infty} d\omega \,\omega^{\beta-1}\cT(s,t)\Bigr\vert_{s=\omega^2,\, t=-z\omega^2}\,.
\end{equation*}

\section{Details on the amplitude techniques}
In this appendix we collect some computations related to the amplitude techniques exploited in the main text. 
\subsection{Froissart-Gribov expansion}
\label{Appx: Details on the Froissart-Gribov expansion}
For the massive case, the Froissart-Gribov expansion is done by going to the center of mass frame and projecting on the little group the eigenfunctions of quadratic Casimir. Given the little group $SO(d-1)$, in general dimension one gets\footnote{For a review of the massive case see~\cite{Correia:2020xtr}}
\begin{equation}
\label{eqpartial wave casimir eq.}
    \Bigl[(1-z^2)^\frac{4-d}{2}\frac{d}{dz}(1-z^2)^\frac{d-2}{2}\frac{d}{dz} + J(J+d-3)\Bigr]P_J^{(d)}(z) =0\,.
\end{equation}
Notice that in this appendix we used $z$ just for simplicity, but it must be regarded as the $z_t$ of the main text. 
If we want to discuss just massless particles, one must suitably modify the above discussion accounting for the different little group, \ie $E(d-2)$. For $d=4$ we can consider the generators $J_3$, $N_1=K_1-J_2$ and $N_2= J_1+K_2$
\begin{equation*}
    [N_1, N_2]=0\,,\qquad [J_3,N_1]=iN_2\,,\qquad [J_3,N_2]=-iN_1\,.
\end{equation*}
However, in order to remove nonphysical continuous spin configuration, we set $C^2 = N_1^2+N_2^2$, \ie the translations of $E(2)$, to zero, thus leaving the little group just the compact $SO(2)$. This is then equivalent to solve the Casimir equation \eqref{eqpartial wave casimir eq.} for $d=3$. It is convenient to write the solutions in terms of  Chebyshev polynomials 
\begin{equation*}
T_J(z)\qquad\text{and}\quad \sqrt{1-z^2}\,U_{J-1}(z)\,,
\end{equation*}
defined for $z\in [-1,1]$. These play the role analogue to $P^{(d)}_J(z)$ in the the massive discussion, \cite{Correia:2020xtr}, so also in the main text we kept the notation $P_J(z)=T_J(z)$. The orthogonality relation is given by
\begin{equation*}
    \int_{-1}^{1}\frac{dz}{\sqrt{1-z^2}} \,T_J(z)T_{J'}(z) = \delta_{J,J'}\Bigr(\frac{\pi}{2} +\frac{\pi}{2}\delta_{J,0}\Bigl)=\delta_{J,J'} \cN_J\,.
\end{equation*}
Moreover, $T_J(z)$ and $U_J(z)$ are related by
\begin{equation*}
    P.V.\int_{-1}^{1}\frac{dw}{\sqrt{1-w^2}} \,T_J(w) \frac{1}{w-z} = \pi\, U_{J-1}(z)\,.
\end{equation*}
Crucially, there exist a relation that allows us to relate $T_J(z)$ with other solutions of the Casimir equation. One of such relation is given by
\begin{equation}
\label{eq from PJ to QJ}
    \frac{1}{2} \int_{-1}^{1}dw\frac{\sqrt{z^2-1}}{\sqrt{1-w^2}}\,\frac{T_J(w)}{z-w} =Q_{J}(z)\,,
\end{equation}
where
\begin{equation}
\label{eq QJ(z) explicit}
    Q_J(z) = \frac{\pi}{2^{1+J}\,z^{J}}\,{}_2 F_1\Bigl(\frac{J}{2},\frac{1+J}{2},1+J;\frac{1}{z^2}\Bigr) = \frac{\pi}{2}\Bigl(1+\sqrt{1-\frac{1}{z^2}}\Bigr)^{-J} z^{-J}\,,
\end{equation}
as also discussed in \cite{Correia:2020xtr}. 

\subsection{Discontinuities and crossing}
\label{app: details on the u- dicontinuity}
Here we collect some additional details concerning crossing symmetry and the relation between the $s$- and $u$-channel cuts.When deforming the contour in \eqref{eq amplitude as pole plus C contour}, in addition to the contribution along the positive real axis in the complex $s$-plane, one must also account for the contribution along the negative real axis. This latter contribution can be reinterpreted as the $u$-channel contribution associated with the physics of on-shell propagation in the $u$-channel.
We therefore rewrite the integral over the discontinuity explicitly in terms of the $u$ variable by a simple change of integration variable
\begin{equation*}
    \frac{1}{2\pi i}\int_{-\infty}^{-t}ds'\frac{\cT(s'+i\epsilon,t)}{s'+i\epsilon-s}+ \frac{1}{2\pi i}\int_{-t}^{-\infty}ds'\frac{\cT(s'-i\epsilon,t)}{s'-i\epsilon-s}=\frac{1}{\pi}\int^{-t}_{-\infty} ds' D_s(\frac{\cT(s',t)}{s'-s})\,.
\end{equation*}
Now for $t$ fixed, we consider $s'=-u'-t$ thus leading to
\begin{equation}
\label{eq integral of discontinuty in u}
    \frac{1}{\pi}\int_0^{\infty}du' \frac{D_u(u',t)}{u'-u}\,.
\end{equation}
We remark that here and in the main text we use the conventional notation 
\begin{equation*}
    \cT(s,t)= \cT(s,t,u)
\end{equation*}
where the last dependence can be neglected due to the constraint $s+t+u=0$. Similarly, we define the crossing symmetric one by
\begin{equation*}
    \cT(u,t)= \cT(u,t,s)\,.
\end{equation*}
Crossing symmetry can be expressed as the invariance of the amplitude under the exchange of channels. In particular, under the interchange of the $s$- and $u$-channels, one has 
\begin{equation*}
    \cT(s,t,u)=\cT(u,t,s)\implies \cT(s,t)=\cT(u,t)\,
\end{equation*}
We this expressions we can rewrite explicitly the discontinuity as
\begin{equation*}
    2i\,D_s(s,t) = \cT(s+i\epsilon,t,u-i\epsilon)-\cT(s-i\epsilon,t,u+i\epsilon)
\end{equation*}
where the $u$ entry gets a contribution too for consistency with the Mandelstam constraint. We see, therefore, that if we explicitly assume the amplitude to be crossing symmetric, we find the relation
\begin{equation*}
    D_s(s,t)=-D_u(u,t)\,,
\end{equation*}
which, when substituted in the expression, it compensate the minus sign of the pole. This contribution is the one  associated with the discontinuity of the $u$-cut. We now rewrite it in terms of the variable $z_t$ where, we recall, that $u=-(z_t+1)t/2$. We thus have
\begin{equation*}
    \cT(s(z_t,t),t)= \cT(z_t,t)\quad \text{and}\quad s(-z_t,t)=u(z_t,t)\implies \cT(u(z_t,t),t):= \cT(-z_t,t)\,
\end{equation*}
so, for example, $\cT(u+i\epsilon,t)=\cT(u(z_t-i\epsilon,t),t)\equiv\cT(-z_t-i\epsilon,t)$. By recollecting all the contributions we get 
\begin{equation}
\label{eq disc o u in zt}
    D_u(u(z_t,t),t)= -D_{z_t}(-z_t,t)\,,
\end{equation} 
where
\begin{equation*}
    D_{z_t}(-z_t,t)=\frac{1}{2i}\Bigl(\cT(-z_t+i\epsilon,t)-\cT(-z_t-i\epsilon,t)\Bigr)\,.
\end{equation*}
In this perspective the different discontinuities are related by the analytic continuation. In particular, we implement crossing by taking different boundary values for the amplitude. For general complex values of $z_t$, by defining $z_t=e^{i\theta}|z_t|$, the different channels can be defined as the different boundary values of the amplitude
\begin{equation}
\label{eq boundaries of amplitude}
    \cT(z_t,t)\Bigr\vert_{\theta=0}=\cT(|z_t|,t)=\cT(s,t)\qquad \text{and}\qquad \cT(z_t,t)\Bigr\vert_{\theta=\pi}=\cT(-|z_t|,t)=\cT(u,t)\,.
\end{equation}
In the discussion, we drop the notation $|z_t|$ leaving it implicit from the definition of the specific contour or integral domain.

Thus, by changing variable in \eqref{eq integral of discontinuty in u} we just get
\begin{equation*}
    \frac{1}{\pi}\int_{-1}^{-\infty}dz_t'\frac{-D_{z_t'}(-z_t',t)}{z_t'-z_t}\,.
\end{equation*}
When we insert this expression on \eqref{eq partial wave coefficients} we get
\begin{equation*}
    \frac{1}{2\cN_J}\int_{-1}^{-\infty}dz_t'\int_{-1}^{1}dz_{t}\,\frac{P_J(z_t)}{\sqrt{1-z_t^2}}\frac{1}{\pi}\,\frac{-D_{z_t'}(-z_t',t)}{z_t'-z_t}\,.
\end{equation*}
By employing \eqref{eq from PJ to QJ}, we obtain 
\begin{equation*}
    \frac{1}{2\pi\cN_J}\int_{-1}^{-\infty}dz_t' \frac{-D_{z_t'}(-z_t',t)2Q_J(z_t')}{\sqrt{z_t'^2-1}}\,.
\end{equation*}
We want now to relate this integral to the $s$-channel one. In order to do so, we need to consider a complex rotation, as discussed in \eqref{eq boundaries of amplitude}. Notice that, in doing so, we get a phase coming from passing the branch cut of the density $(z^2-1)^{-1/2}$. We thus get
\begin{equation*}
     -\frac{1}{\pi\cN_J}\int_{1}^{\infty}dz_t' \frac{D_{z_t}(z_t',t)Q_J(-z_t')}{\sqrt{z_t'^2-1}}\,.
\end{equation*}
If we want to re-express this in terms of the discontinuity of $u$, just by considering \eqref{eq disc o u in zt}, we get $D_{z_t}(z_t,t) = -D_{u}(u(-z_t,t),t)$, thus obtaining
\begin{equation*}
    \frac{1}{\pi\cN_J}\int_1^{\infty} dz_t'\frac{D_u(u(-z_t',t),t)Q_J(-z_t')}{\sqrt{z_t'^2-1}}\,,
\end{equation*}
which will appear on \eqref{eq partial amplitude explicit}.

\subsection{Partial amplitudes at large spin}
\label{app: Details on the large spin partial amplitudes behavior}
In the main text we discussed, for brevity, amplitudes that are crossing symmetric, where contributions coming from the discontinuity appear only for even spins. Here,
we briefly comment why the splitting into definite spin signature must also be used in more general contexts \eg with no crossing symmetry, such as \eqref{eq amplitude as poles plus disc - non-cross-sym}, to properly analytically continue to complex spins. For further details we refer to \cite{Collins:1977jy}. 

We first start by re-expressing the partial amplitudes in $z_t$. Neglecting pole contributions we obtain
\begin{equation}
\label{eq app part wave}
    a_J(t)\supset \frac{1}{\pi\cN_J}\int_{1}^{\infty}dz_t \frac{D_{z_t}(z_t,t)Q_J(z_t)}{\sqrt{z_t^2-1}}+\frac{1}{\pi\cN_J}\int_{-\infty}^{-1}dz_t \frac{D_{z_t}(z_t,t)Q_J(z_t)}{\sqrt{z_t^2-1}}\,.
\end{equation}
As we are interested in performing the analytic continuation on $J$, it is crucial to consider the behavior of such expressions at large $J$. In this regard one must pay particular attention on the domain (in $z_t$) of the two contributions. In particular,  when considering positive $z_t>1$, \ie the $s$-cut, we have that the first contribution grows as
\begin{equation*}
    Q_{J}(z_t)\sim e^{-J\xi(z_t)} \quad\text{with}\quad \xi(z_t)=\log(z_t+\sqrt{z_t^2-1})
\end{equation*}
as in the massive case, with $\xi(z_t)$ real and positive. If we call the integrand of the first contribution \eqref{eq app part wave}, then one has  schematically 
\begin{equation*}
    a_J^{z_t>1}\sim f(z_t,t)e^{-J\xi(z_t)}\,,
\end{equation*}
where $f(z_t,t)$ collects all the remaining terms. This expression converges for $J \to \infty$ and is bounded in the complex directions. When instead we consider the \lhs contributions to the partial amplitudes, \ie when $z_t < -1$, one must take into account the analytic properties of the function $\xi(z_t)$. This part of the partial amplitude would thus grow as
\begin{equation*}
    \xi(z_t)=\xi(\vert z_t\vert)+i\pi\,,
\end{equation*}
related to the other $a_J^{z_t>1}$ by a half rotation in the complex plane. So, the \lhs contribution has the asymptotic behavior of
\begin{equation*}
    a_J^{\,z_t<-1}\sim f(z_t,t) e^{-J\xi(\vert z_t\vert)-i\pi J}\,,
\end{equation*}
which diverges for complex directions, \ie $J\rightarrow i \infty$.
This signals the necessity of first rearranging the partial amplitudes expression before taking the analytic continuation. This is precisely where the definite signature splitting comes to the rescue. By first taking linear combinations of the two contributions, we obtain partial amplitudes that have better behaviour.

So, let us first exploit crossing symmetry to relate the $u$-cut discontinuity to the $s$-one. This leads to the expression

\begin{equation}
\label{eq crossing on aj}
    a_J(t)\supset \frac{1}{\pi\cN_J}\int_{1}^{\infty}dz_t \frac{D_{z_t}(z_t,t)Q_J(z_t)}{\sqrt{z_t^2-1}}+\frac{1}{\pi\cN_J}\int_{1}^{\infty}dz_t' \frac{D_{z_t}(-z_t,t)(-1)^{-J}Q_J(z_t)}{\sqrt{z_t^2-1}}\,.
\end{equation}

We can therefore first define definite signature contributions
\begin{equation*}
    D^{\pm}_{z_t}(z_t,t)= D_{z_t}(z_t,t)\pm D_{z_t}(-z_t,t)\,,
\end{equation*}
and the corresponding coefficients
\begin{equation*}
    a^{\pm}_J(t)\supset \frac{1}{\pi\cN_J}\int_{1}^{\infty}dz_t \frac{D^\pm_{z_t}(z_t,t)Q_J(z_t)}{\sqrt{z_t^2-1}}\,.
\end{equation*}
From these expressions, we can consistently analytically continue in the complex $J$ plane. Moreover, when crossing symmetry is enforced by
\begin{equation*}
     D_{z_t}(z_t,t)= D_{z_t}(-z_t,t)\,,
\end{equation*}
one finds only even contributions appearing, as discussed in section \ref{sec: Regge limit}.

\subsection{Partial amplitudes in the complex $J$ plane}
\label{app: regge pole}
Here we briefly provide additional details on the properties and motivation underlying the partial-wave representation. For further details we see \cite{Collins:1977jy}.

In particular, we note that the expressions for $a_J^{\pm}(t)$ are well defined only when the corresponding integrals converge. From \eqref{eq QJ(z) explicit}, we know that for large $z_t$
\begin{equation*}
(z_t^2-1)^{-1/2} Q_J(z_t) \sim z_t^{-1-J}.
\end{equation*}
Assuming in full generality that $D_s(s(z_t),t)\sim z_t^{N(t)}$, the integral converges for sufficiently large spin provided
\begin{equation*}
\Re(J) > N(t),
\end{equation*}
while it diverges otherwise. 
The first important observation is that, given convergence as $|J|\to\infty$, the contribution from the arcs at infinity in the complex integral in $J$ can be safely neglected. We may then study what happens as $J$ is lowered towards the critical value $\Re(J)\simeq\alpha(t)$, where we identify $N(t)$ in the standard way.

In this regime, for $z_t\gg1$, the integral can be approximated as
\begin{equation*}
\int_{z_t^\star}^{\infty}dz_t, z_t^{\alpha(t)}z_t^{-1-J}=-\frac{e^{(\alpha(t)-J)\log z_t^\star}}{\alpha(t)-J},
\end{equation*}
where $z_t^\star$ is some cutoff to approximate the integral to high values of $z_t$. We see that the partial amplitude exhibits a pole at $J=\alpha(t)$ in the complex $J$-plane. This is one of the motivation for which one typically assumes the existence of the Regge pole. On the complex plane there could be other singularities and branch cuts. However, we assume the Regge pole to be the first singularity we find as we lower $J$, with all the other possible poles to give a subleading contributions or branch cuts that can be avoided by key-hole contour modifications in \eqref{eq. leading regge pole plus contour}. 
Crucially, notice that the Regge pole comes from the discontinuity part of the partial amplitude. These contributions are not present when we discuss theories at tree-level, so we may regard these as loop contributions. Moreover, for crossing symmetric amplitude, since only $a^{+}_J(t)$ has the explicit dependence on the integral, only the even signature contributions will feature this poles. The only exception to all this arguments is when the theory actually has an infinite set of poles. This is the case of stringy amplitudes which exhibits Regge trajectories already at tree-level.

\section{$SL(2,\mathbb{C})$ CFT }
\label{app: sl2 CFT}
In this appendix we briefly introduce the notations and we review the main features of the $SL(2,\bC)$ CFT conformal block expansion.
\subsection{Review of the $SL(2,\mathbb{R})$ algebra}
We first begin from the real $SL(2,\mathbb{R})$ and the complex case will simply follow. The algebra is given by
\begin{equation*}
    [L_n,L_m]=(n-m)L_{n+m}\,.
\end{equation*}
It allows to identify the infinite dimensional representation for the generator 
\begin{equation*}
    L_n=-h(n+1)z^n -z^{n+1}\partial_z
\end{equation*}
that satisfy the algebra. From this, we find the quadratic Casimir  of the global part to be given by
\begin{equation*}
    \cC_2 = L_0^2 -\frac{1}{2}\Bigl( L_{-1}L_{1}+L_1 L_{-1} \Bigr)\,,
\end{equation*}
such that, when acting on test functions, it gives
\begin{equation*}
    \cC_2 \,f(z;h)=h(h-1)f(z;h)\,.
\end{equation*}

In order to obtain the conformal blocks, as usual, we insert the quadratic Casimir in a four-point function. In this respect, in order to distinguish the action on different insertions, we redefine 
\begin{equation*}
    L_{n\,i}=-h_i(n+1)z_i^n -z_i^{n+1}\partial_{z_i}\,,
\end{equation*}
which acts on the respective $\cO_{h_i}(z_i)$. By defining 
\begin{equation*}
    \mathrm{L}_n = L_{n\, 1}+L_{n\, 2}\,,
\end{equation*}
we obtain
\begin{equation*}
    \mathrm{C}_{1,2} = \mathrm{L}_0^2 -\frac{1}{2}\Bigl( \mathrm{L}_{-1}\mathrm{L}_{1}+\mathrm{L}_1 \mathrm{L}_{-1} \Bigr)\,,
\end{equation*}
which, acting on the first and second insertions of the correlator, allows to obtain the differential conformal Casimir equation.
We fix the conformal frame with  $z_4 =\infty$ so that $z=\frac{z_{12}}{z_{13}}$ and the four-point function reduces to
\begin{equation*}
    \vev{\cO_{h_1}(z_1)\cO_{h_2}(z_2)\cO_{h_3}(z_3)\cO_{h_4}(z_4)}\sim \frac{1}{(z_{12})^{h_1+h_2}(-z_4)^{h_3+h_4}}\Bigl(\frac{-z_4}{z_{13}}\Bigr)^{h_3-h_4}G(\frac{z_{12}}{z_{13}})\,.
\end{equation*}
Expanding $G(z)$ in conformal blocks $k_{h}^{h_i}(z)$ and acting with $\mathrm{C}_{1,2}$ we find the differential equation 
\begin{equation*}
    \Bigl(z^2(1-z)\partial_z^2-(1-h_{12}+h_{34})z^2 \partial_z+h_{12}h_{34}z\Bigr)k_{h}^{h_i}(z) =h(h-1)\,k_{h}^{h_i}(z) \,,
\end{equation*}
which is solved by
\begin{align}
    \tilde k_{h}(z)&= (-z)^h\,_2F_1(h-h_{12},h+h_{34},2h,z)\,,\\
    \tilde k_{1-h}(z)&= (-z)^{1-h}\,_2F_1(1-h-h_{12},1-h+h_{34},2-2h,z)\,.
\end{align}
The differential equation does not fix an overall constant. We fix it requiring that the behavior of $k_h(z)$ near $z\sim0$ is $k_h(z)\sim z^h$ which is the leading contribution of the exchanged primaries in the s-channel. So, we will consider
\begin{align}
     k_{h}(z)&= z^h\,_2F_1(h-h_{12},h+h_{34},2h,z)\,,\\
     k_{1-h}(z)&= z^{1-h}\,_2F_1(1-h-h_{12},1-h+h_{34},2-2h,z)\,.
\end{align}

\subsection{$SL(2,\mathbb{C})$ blocks}
Given that $\sl(2,\mathbb{C})\simeq \sl(2,\mathbb{R})\times \sl(2,\mathbb{R})$ we can build the conformal blocks by taking the products of the one above, now with $z\in \mathbb{C}$ and $\zb = z^*$. 
By taking $h = \frac{\Delta+J}{2}$ and $\bar h = \frac{\Delta-J}{2}$, the conformal blocks for $SL(2,\mathbb{C})$ are given by the real combination
\begin{equation*}
    G_{\Delta,J}(z,\zb) =\frac{1}{1+\delta_{J,0}}\Bigl( k_{\Delta+J}(z)k_{\Delta-J}(\zb) +k_{\Delta-J}(z)k_{\Delta+J}(\zb)\Bigr)\,,
\end{equation*}
where
\begin{equation*}
     k_{\Delta+J}(z)= z^{\frac{\Delta+J}{2}}\,_2F_1\Bigl(\frac{\Delta+J}{2}-\frac{\Delta_{12}+J_{12}}{2},\frac{\Delta+J}{2}+\frac{\Delta_{34}+J_{34}}{2},\Delta+J;z\Bigr)\,.
\end{equation*}
This follows from the implicit assumption that $h,\bar h\in\bR$, so also $\Delta,J\notin \bC$. However, we will be interested in consider cases with $h,\bar h\in\bC$ and  generically with $\bar h\neq h^*$, in order to make contact with the celestial CFT discussion. Indeed, as known \cite{Pasterski:2017kqt,Lam:2017ofc}, the basis for celestial primaries operators is given by operator in the principal series $\Delta=1+i\lambda$, with $\lambda\in\bR$. 
We thus consider the CFT conformal partial waves expansion \cite{Dolan:2011dv,Hogervorst:2013kva,Caron-Huot:2017vep,Simmons-Duffin:2012juh}
\begin{equation}
\label{eq harmonic function}
    F_{\Delta,J}(z,\zb) =\frac{1}{2}\Bigl( G_{\Delta,J}(z,\zb)+  \frac{K_{2-\Delta,J}}{K_{\Delta,J}} G_{2-\Delta,J}(z,\bar z) \Bigr)\,,
\end{equation}
where 
\begin{equation}
\label{eq K gamma combination}
    K_{\Delta,J}=\frac{\Gamma(\frac{\Delta+J+\Delta_{12}}{2}){\Gamma(\frac{\Delta+J-\Delta_{12}}{2})\Gamma(\frac{\Delta+J+\Delta_{34}}{2})\Gamma(\frac{\Delta+J-\Delta_{34}}{2})}}{2\pi^2 \Gamma(\Delta+J-1)\Gamma(\Delta+J)}\,.
\end{equation}

This expression allows to expand the cross-ratio function as
\begin{equation}
\label{eq harmonic function expansion}
   f_{\Delta_i}(z,\zb)= \sum_{J=0}^{\infty}\int_{1-i\infty}^{1+i\infty}\frac{d\Delta}{2\pi i}c(\Delta,J) F_{\Delta,J}(z,\zb)\,,
\end{equation}
where the $c(\Delta,J)$ residues encodes the OPE information. Notice that given the property 
\begin{equation*}
    \frac{c(\Delta,J)}{K_{\Delta,J}}=\frac{c(2-\Delta,J)}{K_{2-\Delta,J}}\,,
\end{equation*}
we simply obtain that
\begin{equation}
\label{eq principal series expansion in conformal blocks}
    f_{\Delta_i}(z,\zb)=\sum_{J=0}^{\infty}\int_{1-i\infty}^{1+i\infty}\frac{d\Delta}{2\pi i}c(\Delta,J) G_{\Delta,J}(z,\zb)\,.
\end{equation}

\section{Evaluating the monodromies}
\label{app: monodromies}
We collect here some dettails on the computations for the monodromies of the conformal blocks and hypergeometric functions.
\subsection{Monodromies of the hypergeometric functions}
The hypergeometric functions have three branching points at $z=0,1$ and $z=\infty$. Since these will play an important role when discussing the Lorentzian setups, we start with a systematic analysis of these special points and their monodromies.\\
First of all, we start from the hypergeometric functions as solutions of the differential equation
\begin{equation}
\label{eqhypergeom differential equation}
  \Bigr(  z(1-z)\partial_z^2 +[c-(a+b+1)z]\partial_z  -a b z\Bigl)f(z) =0\,.
\end{equation}
The solutions are
\begin{equation*}
    f(z) = c_1 \,_2F_1(a,b,c,z) + c_2 (-z)^{1-c}\,_2F_1(1+a-c,1+b-c,2-c,z) = c_1 \varphi_1(z)+c_2 \varphi_2(z)\,.
\end{equation*}
These solutions form  a basis for the monodromy matrix $M_0$, where the path is considered around zero. In order to compute the entries we can either exploit the series expansion for $\abs{z}<1$ or by directly solving the differential equation with the phase included. If we consider $\phi=(\varphi_1,\varphi_2)$ the monodromy matrix is simply  $M_0=\diag (1 , e^{2\pi i (1-c)})$ or, equivalently
\begin{equation*}
    M_0=\diag (1 , e^{-2\pi i c})\,.
\end{equation*}
We may now consider the monodromy around $z=1$, denoted $M_1$. Given that the basis $\phi$ will not diagonalize $M_1$, we can first solve \eqref{eqhypergeom differential equation} for $w=1-z$ around $w=0$. The solutions to 
\begin{equation}
\label{eqhypergeom differential equation w}
  \Bigr(  w(1-w)\partial_w^2 -[c-(a+b+1)(1-w)]\partial_w  -a b (1-w)\Bigl)f(w) =0\,,
\end{equation}
are the generic linear combinations of
\begin{align}
    f(w) &= c_1 \,_2F_1(a,b,1+a+b-c,w) + c_2 (-w)^{c-a-b}\,_2F_1(c-a,c-b,1-a-b+c,w)\nonumber\\
    &= c_1 \psi_1(w)+c_2 \psi_2(w)\nonumber\,.
\end{align}
Notice that these solutions are not simply $\varphi(z)\rightarrow\varphi(1-z)$.
In this basis, \ie $(\psi_1,\psi_2)$, we can easily consider the monodormy around $z=1$ \ie $z\rightarrow 1- e^{2\pi i}(1-z)$. This gives $M_1=\diag(1, e^{2\pi i(c-a-b)})$.\\
We will not use it, but for reference the last monodromy around $z=\infty$ has as basis
\begin{align}
    \varphi_1 &= (-w)^a \,_2F_1(a,1+a-c,1-a+b,w)\,,\nonumber\\
    \varphi_2 &= (-w)^b \,_2F_1(b,1+b-c,1+a+b,w)\,,\nonumber
\end{align}
with $w=1/z$. Now, in this base, the diagonal monodromy matrix is
\begin{equation*}
    M_\infty = \diag(e^{2\pi i a} , e^{2\pi i b})\,.
\end{equation*}

We are mainly interested in relating the monodromy in one basis to the one in other basis. In this perspective we first consider the following identity
\begin{align}
     \,_2F_1(a,b,c,z) = &\frac{\Gamma(c)\Gamma(c-a-b)}{\Gamma(c-a)\Gamma(c-b)} \,_2F_1(a,b,a+b+1-c,1-z)+\nonumber\\
     &+\frac{\Gamma(c)\Gamma(a+b-c)}{\Gamma(a)\Gamma(b)} (1-z)^{c-a-b}\,_2F_1(c-a,c-b,1+c-a-b,1-z)\,.
\end{align}
For brevity we define the two coefficients
\begin{equation*}
    \gamma_1 = \frac{\Gamma(c)\Gamma(c-a-b)}{\Gamma(c-a)\Gamma(c-b)}\,,\qquad \gamma_2 = \frac{\Gamma(c)\Gamma(a+b-c)}{\Gamma(a)\Gamma(b)}\,,
\end{equation*}
in order to write the expression more compactly as\footnote{Notice that the relation is given, rather than $\psi_2$, in terms of $(-1)^{c-a-b}\psi_2$. This sign can be either absorbed in the definition of $\psi_2$ or in the factor $\gamma_2$. In the following we just redefined $\psi_2$ with this factor}
\begin{equation*}
    \varphi_1(z)= \gamma_1\,\psi_1(1-z) + \gamma_2\,\psi_2(1-z)\,.
\end{equation*}
We can consider the action of $M_1$ on $\varphi_1$ by exploiting the expansion of the latter in the basis in which $M_1$ is diagonal. This gives
\begin{equation*}
    M_1\varphi_1(z)= \gamma_1\,\psi_1(1-z) + \gamma_2\,e^{2\pi i(a+b-c)}\psi_2(1-z) = \varphi_1(z) + \gamma_2\Bigl( e^{2\pi i(a+b-c)}-1\Bigr)\, \psi_2(z)\,,
\end{equation*}
which explicitly means
\begin{multline*}
    M_1[\,_2F_1(a,b,c,z)] = \,_2F_1(a,b,c,z) +\\
    +\Bigl( e^{2\pi i(c-a-b)}-1\Bigr)\frac{\Gamma(c)\Gamma(a+b-c)}{\Gamma(a)\Gamma(b)} (1-z)^{c-a-b}\,_2F_1(c-a,c-b,1-a-b+c,1-z)\,
\end{multline*}
By exploiting the relation 
\begin{equation*}
    \Gamma(1+c-a-b)\,\Gamma(a+b-c) = \frac{\pi}{\sin(\pi(a+b-c))}\,,
\end{equation*}
we can rewrite the expression as
\begin{multline}
    M_1[\,_2F_1(a,b,c,z)] = \,_2F_1(a,b,c,z) +\\
    - \frac{2\pi i\,\Gamma(c)\,e^{\pi i(c-a-b)}}{\Gamma(a)\Gamma(b)\Gamma(1+c-a-b)} (1-z)^{c-a-b}\,_2F_1(c-a,c-b,1-a-b+c,1-z)\,.\label{eq monodromy of the hyper}
\end{multline}
Notice, also, that given the triviality of the monodromy of $\,_2F_1(a,b,a+b,z)$ around $z=0$, this is also equivalent to the double monoromy around both $z=0$ and then $z=1$ so
\begin{equation*}
  M_1[ M_0[\,_2F_1(a,b,a+b,z)]]= M_1[\,_2F_1(a,b,a+b,z)]\,.
\end{equation*}

\subsection{Monodromies of the conformal blocks}
\label{app sub: Monodromies of the conformal blocks}
We can now analyze the monodromies of the conformal blocks exploiting the expressions derived in the subsection above. Our focus will be on the monodromy around $z=1$, while keeping the ratio $z/\bar{z}$ fixed. As we shall see, the limit $z \ll 1$ corresponds to the Regge regime of the CFT. From \eqref{eq monodromy of the hyper} and the expression for the conformal blocks \eqref{eq harmonic function}, the direct evaluation shows that the leading contribution in $z$ is given by
\begin{equation*}
    G_{\Delta,J}^{\circlearrowleft}(z,\zb)\approx  \frac{2i\pi e^{\frac{i\pi (\Delta_{12}-\Delta_{34}+J_{12}+J_{34})}{2}}\,z^{1-J}\Bigl(\frac{z}{\zb}\Bigr)^{\frac{J-\Delta}{2}}\frac{\Gamma^2(\Delta+J)}{(1-J-\Delta)}}{\Gamma(\frac{\Delta+J+\Delta_{12}+J_{12}}{2}){\Gamma(\frac{\Delta+J-\Delta_{12}-J_{12}}{2})\Gamma(\frac{\Delta+J+\Delta_{34}+J_{34}}{2})\Gamma(\frac{\Delta+J-\Delta_{34}-J_{34}}{2})}}
\end{equation*}
Evaluating for external spinless particles one simply gets
\begin{equation*}
    G_{\Delta,J}^{\circlearrowleft}(z,\zb)\approx  \frac{2i\pi e^{\frac{i\pi (\Delta_{12}-\Delta_{34})}{2}}\,z^{1-J}\Bigl(\frac{z}{\zb}\Bigr)^{\frac{J-\Delta}{2}}\frac{\Gamma^2(\Delta+J)}{(1-J-\Delta)}}{\Gamma(\frac{\Delta+J+\Delta_{12}}{2}){\Gamma(\frac{\Delta+J-\Delta_{12}}{2})\Gamma(\frac{\Delta+J+\Delta_{34}}{2})\Gamma(\frac{\Delta+J-\Delta_{34}}{2})}}
\end{equation*}
Here we observe the characteristic $z^{1-J}$ behavior in the Regge limit. This contribution arises from analytically continuing $z$ around the branch point. For $z \ll 1$, it provides the leading term in the partial wave expansion, precisely because of the divergent factor $z^{1-J}$.
It can be rewritten by noticing that $\Gamma(\Delta+J-1)\Gamma(\Delta+J)=\Gamma(\Delta+J)^2/(\Delta+J-1)$ thus leading to 
\begin{equation*}
    G_{\Delta,J}^{\circlearrowleft}(z,\zb)\approx  \frac{-2i\pi \,e^{\frac{i\pi (\Delta_{12}-\Delta_{34})}{2}}\,z^{1-J}\Bigl(\frac{z}{\zb}\Bigr)^{\frac{J-\Delta}{2}}\Gamma(\Delta+J-1)\Gamma(\Delta+J)}{\Gamma(\frac{\Delta+J+\Delta_{12}}{2}){\Gamma(\frac{\Delta+J-\Delta_{12}}{2})\Gamma(\frac{\Delta+J+\Delta_{34}}{2})\Gamma(\frac{\Delta+J-\Delta_{34}}{2})}}
\end{equation*}
By exploiting \eqref{eq K gamma combination} we can rewrite it in the more compact way
\begin{equation}
\label{eq monodromy of the block z=1}
     G_{\Delta,J}^{\circlearrowleft}(z,\zb)\approx  \frac{-i\, e^{\frac{i\pi (\Delta_{12}-\Delta_{34})}{2}}}{\pi\, K_{\Delta,J}}z^{1-J}\Bigl(\frac{z}{\zb}\Bigr)^{\frac{J-\Delta}{2}}\,.
\end{equation}
Given the fact that block is symmetric under the real $z$ and $\zb$, \ie $G_{\Delta,J}(z,\zb)=G_{\Delta,J}(\zb,z)$, the evaluation of the monodromy in both variables will lead to an analogous result where the two are swapped $z \leftrightarrow\zb$,
\begin{equation}
\label{eq monodromy of the block zbar=1}
   G_{\Delta,J}^{\circlearrowleft}(z,\zb)\approx  \frac{-i\, e^{\frac{i\pi (\Delta_{12}-\Delta_{34})}{2}}}{\pi\, K_{\Delta,J}}\zb^{1-J}\Bigl(\frac{\zb}{z}\Bigr)^{\frac{J-\Delta}{2}}\,.
\end{equation}

This is true provided we take the same path. Notice, that while the structure is similar, the causal configurations associated with the two expressions are different. As known, \cite{Caron-Huot:2017vep,Kravchuk:2018htv}, the characteristic behavior for small values of $z$ ($\zb$) can be obtained by the expansion of a block $G_{1-J,1-\Delta}(z,\zb)$. This contribution is associated with the exchanged of a different kind of operators called light-ray operators.

\section{Dealing with the distributions}
\label{app: Dealing with the distributions}

We want to perform the same analytic manipulations on both sides of the expression \eqref{eq reduced correlator f of the t-kinematics} where in the \lhs we perform the CFT procedure and in the \rhs the celestial one. However, the first obstruction of applying directly this approach is the non-holomorphicity of the delta function. We therefore begin by introducing a proper prescription to actually consider analytic manipulations, such as monodromies, in this peculiar scenario. We introduce these arguments with a brief example. Let us take then a generic distribution $f$ acting on the test function space 
\begin{equation}
\label{eq distrib}
    \langle f\,, \phi\rangle = \int dz d\zb f(z,\zb) \phi(z,\zb)\,.
\end{equation}
where $\phi\in\cD(\bC)$ \ie smooth single-valued test functions on $\bC$. We consider now a particular distribution that we define
\begin{equation}\label{eq sing distr}
    f(z,\zb) = \delta(z-\zb)\, h(z)\,.
\end{equation}
From this, we can evaluate explicitly \eqref{eq distrib} thus obtaining
\begin{equation*}
     \langle f\,, \phi\rangle = \int_0^\infty dr \int_0^{2\pi} d\theta \,\delta(\theta)g(r,\theta)\phi(r,\theta) = \int_0^{\infty}dr\, h(r,0) \phi(r,0)\,.
\end{equation*}
From this perspective, we see that considering the distribution $f$ acting on $\mathcal{D}(\mathbb{C})$ is equivalent to the action of a new distribution $\mathrm{f}$, which we relate to $h(z)$ by $= h(r,0) \equiv \mathrm{h}(r)$.
The distribution $\text{f}$ is now considered as acting on the space of test functions $\cD(\bR^+)\subset \cD(\bC)$. We thus have the equivalence between
\begin{equation}\label{eq distrib eq}
      \langle f\,, \phi\rangle \equiv  \langle \text{f} \,, \varphi\rangle\,,
\end{equation}
where now
\begin{equation*}
     \langle \text{f} \,, \varphi\rangle = \int_0^{\infty}dr\, \text{h}(r) \varphi(r)\,.
\end{equation*}
Essentially, \eqref{eq distrib eq} tells us that since all the information of the first distribution is supported on the locus $z=\zb$, we may equivalently define a restriction homomorphism $\cD(\bC)\rightarrow\cD(\bR^+)$.

This line of reasoning allows us to introduce a scheme that can be exploited to define a different--yet related--distribution, which no longer suffers from issues related to analyticity. In practice, this amounts to studying the analytic structure on top of the delta-function support.
A similar discussion also provides a consistent prescription for performing a Wick rotation, mapping the celestial CFT to a Lorentzian configuration.

Based on these considerations, we can now make sense on what we mean by taking the monodromy of a singular distribution such as \eqref{eq sing distr}. Indeed, as a consequence of the equivalence relation \eqref{eq distrib eq}, we define the monodromies of $f(z,\zb)$ in terms of those of $\text{f}(r)$. Thus, with a slight abuse of notation we define
\begin{equation*}
    M_{z=1}[f ]\equiv M_{z=1}[\text{f} ]\,,\qquad  M_{\zb=1}[f ]\equiv M_{\zb=1}[\text{f} ]\,.
\end{equation*}
or, in other terms
\begin{equation*}
    M_{z=1}[f(z,\zb) ]=  \delta(z-\zb)\,M_{z=1}[h(z)]\,,\qquad  M_{\zb=1}[f(z,\zb) ]=  \delta(z-\zb)\,M_{z=1}[h(z)]\,,
\end{equation*}
where $M_{z=1}$ ($M_{\zb=1}$) takes the monodromy on $z$ ($\zb$) around the given value. Notice that an equivalent result can be obtained from the  Sokhotski-Plemelj formula where, after taking linear combinations, one has
\begin{equation*}
   2\pi i  \delta(z-\zb)= \lim_{\epsilon\rightarrow0^+}\Bigl(\frac{1}{z-\zb-i\epsilon}-\frac{1}{z-\zb+i\epsilon}\Bigr)\,.
\end{equation*}
This is similar to the case discussed here, where one instead assumes that the monodromy commutes with the limit, and so the monodromies would act only on the $h(z)$ part.

\bibliographystyle{JHEP}
\nocite{*}
\bibliography{crl.bib}

\end{document}